\documentclass[aps,prd,11pt,twoside,tightenlines,nofootinbib,showpacs,preprint,superscriptaddress]{revtex4-1}
\usepackage[colorlinks=true]{hyperref}
\usepackage{xcolor}
\hypersetup{colorlinks=true, citecolor=blue, urlcolor=blue, linkcolor=blue}
\usepackage{amsmath,amssymb}
\usepackage[final]{graphicx}
\usepackage{subcaption}
\usepackage{float}
\usepackage[toc,page]{appendix}

\begin{document}

\title{Scalar resonances in the final state interactions of the decays $D^{0} \rightarrow \pi^{0}\pi^{0}\pi^{0}, \pi^{0}\pi^{0}\eta, \pi^{0}\eta\eta$}

\author{Zhong-Yu Wang}
\email{zhongyuwang@foxmail.com}
\affiliation{School of Physics and Electronics, Hunan Key Laboratory of Nanophotonics and Devices, Central South University, Changsha 410083, China}

\author{Hiwa A. Ahmed}
\email{hiwa.ahmed@charmouniversity.org}
\affiliation{Medical Physics Department, College of Medical and Applied Science, Charmo University, Chamchamal, Sulaymaniyah 46023, Iraq}
\affiliation{School of Nuclear Science and Technology, University of Chinese Academy of Sciences, Beijing 100049, China}

\author{C. W. Xiao}
\email{xiaochw@csu.edu.cn}\thanks{ (corresponding author)}
\affiliation{School of Physics and Electronics, Hunan Key Laboratory of Nanophotonics and Devices, Central South University, Changsha 410083, China}

\date{\today}

\begin{abstract}

We investigate the scalar resonances in the processes of $D^{0} \rightarrow \pi^{0}\pi^{0}\pi^{0}, \pi^{0}\pi^{0}\eta, \pi^{0}\eta\eta$ decays based on the chiral unitary approach for the final state interaction. We start from singly Cabibbo-suppressed production diagrams which provide a primary quark pair to hadronize two pseudoscalar mesons in $D^{0}$ decays. The resonances $f_{0}(500)$, $f_{0}(980)$ and $a_{0}(980)$ are dynamically produced from the final state interactions of the meson pairs. In our results, the experimental data for the $\pi^{0}\eta$ invariant mass spectrum of the $D^{0} \rightarrow \pi^{0}\eta\eta$ decay can be described well. We also make the predictions for the $\pi^{0}\pi^{0}$ invariant mass spectrum of the $D^{0} \rightarrow \pi^{0}\pi^{0}\pi^{0}$, where the $f_{0}(980)$ can be found, and for the $\pi^{0}\pi^{0}$, $\pi^{0}\eta$ invariant mass spectra of the $D^{0} \rightarrow \pi^{0}\pi^{0}\eta$, where the $f_{0}(500)$, $f_{0}(980)$ and $a_{0}(980)$ appear. Furthermore, the branching ratios of each decay channel are predicted. We expect more accurate measurements of these decays to better understand the nature of the states $f_{0}(500)$, $f_{0}(980)$ and $a_{0}(980)$.

\end{abstract}

\maketitle

\section{Introduction}

The three-body charmed meson decays have become important sources for investigating the nature of low-lying scalar resonances. Due to the existence of three mesons in the final states, a large number of scalar resonances are produced in these processes. Recently, many experiments have reported the three-meson decay channels of $D$ mesons. The LHCb Collaboration performed Dalitz plot analysis of the doubly Cabibbo-suppressed decay $D^{+} \rightarrow K^{-}K^{+}K^{+}$ for the first time in Ref. \cite{LHCb:2019tdw}, where the structures of $\phi(1020)$, $f_{0}(980)$,  $f_{0}(1370)$ and $a_{0}(980)$ states in this decay process were studied. The BESIII Collaboration observed the singly Cabibbo-suppressed decay $D^{+}\rightarrow\eta\eta\pi^{+}$ and accurately measured the branching fractions of $D^{+}\rightarrow\eta\pi^{+}\pi^{0}$ and $D^{0}\rightarrow\eta\pi^{+}\pi^{-}$ in Ref. \cite{BESIII:2019xhl}, which offered an opportunity to investigate the decays $D\rightarrow\rho\eta, a_{0}(980)\pi, a_{0}(980)\eta$.  In Ref. \cite{Belle:2020fbd}, the decay of $D^{0} \rightarrow K^{-}\pi^{+}\eta$ was investigated via Dalitz plot analysis by Belle Collaboration, where the $\bar{K}^{*}(892)^{0}$, $a_{0}(980)^{+}$, $a_{2}(1320)^{+}$, etc, were found, and the ratios of branching fractions in different decay channels were measured. In Ref. \cite{BESIII:2020ctr}, the BESIII Collaboration had reported the amplitude analysis results and the most precise branching fraction measurement of $D_{s}^{+} \rightarrow K^{+}K^{-}\pi^{+}$, which were consistent with those obtained in previous experiments \cite{E687:1995jyc, CLEO:2009nuz}. Moreover, one can find plenty of the processes of $D$ meson decaying into three pseudoscalar mesons in Particle Data Group (PDG) \cite{Zyla:2020pdg}. In PDG, one can find that most of the final states of $D$ meson three-body decays contain charged mesons, but there are a few processes having three neutral particles in the final states. For the unique $D$ decay with three neutral pseudoscalar mesons in the final states, the first search for the decay $D^{0} \rightarrow \pi^{0}\pi^{0}\pi^{0}$ was in 2006 by CLEO Collaboration in Ref. \cite{CLEO:2005mti}, where the “single tag” method was used to obtain a branching fraction upper limit of $3.5 \times 10^{-4}$ at the $90\%$ confidence level. The decays of $D^{0} \rightarrow \pi^{0}\pi^{0}\pi^{0}, \pi^{0}\pi^{0}\eta, \pi^{0}\eta\eta, \eta\eta\eta$ were investigated by the BESIII Collaboration in Ref. \cite{BESIII:2018hui}. The corresponding branching fractions were measured to be $\mathcal{B}\left(D^{0} \rightarrow \pi^{0}\pi^{0}\pi^{0}\right)=(2.0 \pm 0.4 \pm 0.3) \times 10^{-4}$, $\mathcal{B}\left(D^{0} \rightarrow \pi^{0}\pi^{0}\eta\right)=(3.8 \pm 1.1 \pm 0.7) \times 10^{-4}$ and $\mathcal{B}\left(D^{0} \rightarrow \pi^{0}\eta\eta\right)=(7.3 \pm 1.6 \pm 1.5) \times 10^{-4}$, respectively. The $D^{0} \rightarrow \eta\eta\eta$ signal was not observed and the upper limit on its decay branching fraction was $\mathcal{B}\left(D^{0} \rightarrow \eta \eta \eta\right)<1.3 \times 10^{-4}$ at the $90\%$ confidence level. The $a_{0}(980)$ was found in the $\pi^{0}\eta$ invariant mass distribution of $D^{0} \rightarrow \pi^{0}\eta\eta$ decay. These experimental measurements provided an opportunity for studying the contribution of scalar resonances. 

It is challenging to study the three-body decay of $D$ meson theoretically. As done in Ref. \cite{El-Bennich:2006nld} for the scalar form factors of the $D_{(s)} \rightarrow f_{0}(980)$ transition using a covariant quark model, Ref. \cite{El-Bennich:2008rkp} studied the properties of the $f_{0}(980)$ resonance in the $D_{(s)} \rightarrow f_{0}(980)\;\pi/K$ decays and made predictions for the $B_{(s)} \rightarrow f_{0}(980)\;\pi/K$ decays. In Ref. \cite{Cheng:2021yrn}, the $CP$ asymmetries in three-body $D^{0} \rightarrow K^{+}K^{-}\pi^{0}$, $D^{0} \rightarrow \pi^{+}\pi^{-}\pi^{0}$, $D^{+} \rightarrow K^{+}K_{S}\pi^{0}$ and $D_{s}^{+} \rightarrow K^{0}\pi^{+}\pi^{0}$ decays were analyzed through intermediate vector resonances within the framework of topological amplitude approach for tree amplitudes and the QCD factorization approach for penguin amplitudes. The three-body decay processes $D^{0} \rightarrow \pi^{0}\pi^{0}\pi^{0}$ and $D^{0} \rightarrow \pi^{0}\pi^{0}\eta$, and the other decay modes were researched in Ref. \cite{Gershon:2015xra}, where the decay width difference between the two physical eigenstates of the $D^{0}-\bar{D}^{0}$ system were studied. With the analysis of the first measurement of $D^0$ and $D^+$ semileptonic decays $D^0 \to a_0(980)^- e^+ \nu_e$, $D^+ \to a_0(980)^0 e^+ \nu_e$ by BESIII collaboration~\cite{BESIII:2018sjg}, Ref.~\cite{Achasov:2018grq} found no constituent two-quark component in the $a_0(980)$ wave function, where its four-quark production in the semileptonic decays $D\to \eta\pi e^+ \nu_e$ was investigated in a further work of~\cite{Achasov:2021dvt}. Performing the analysis of semileptonic decays $D \to \pi^+ \pi^- e^+ \nu_e$, $D_s \to \pi^+ \pi^- e^+ \nu_e$ from the BESIII and CLEO data~\cite{CLEO:2009ugx,BESIII:2018qmf}, Ref.~\cite{Achasov:2020qfx} supported the interpretation of four-quark nature for the $f_{0}(500)$ and $f_{0}(980)$ resonances, which were also discussed recently in detail in Ref.~\cite{Achasov:2020aun} for their four-quark nature based on the BESIII data of the decay $J/\psi \to \gamma \pi^0 \pi^0$~\cite{BESIII:2015rug}. Moreover, more discussions on the four-quark nature of $f_{0}(500)$, $f_{0}(980)$ and $a_0(980)$ states can be referred to Refs.~\cite{Achasov:1987ts,Achasov:2003cn,Achasov:2009ee,Achasov:2017zhy,Achasov:2019ywj}.
Taking into account the final state interactions with the chiral unitary approach (ChUA) \cite{Oller:1997ti, Oset:1997it, Oller:1997ng, Kaiser:1998fi, Oller:2000fj, Oset:2008qh}, the $D^{0} \rightarrow \bar{K}^{0}\pi^{+}\pi^{-}$ and $D^{0} \rightarrow \bar{K}^{0}\pi^{0}\eta$ decays were investigated in Ref. \cite{Xie:2014tma}, where the contributions of the low-lying scalar resonances $f_{0}(500)$, $f_{0}(980)$ in $\pi^{+}\pi^{-}$ components and $a_{0}(980)$ in $\pi^{0}\eta$ were reproduced and a ratio obtained with their contributions to the branching ratios was in good agreement with experiment. Ref. \cite{Dias:2016gou} studied the $f_{0}(980)$ production in the $D_{s}^{+} \rightarrow \pi^{+}\pi^{+}\pi^{-}$ and $D_{s}^{+} \rightarrow \pi^{+}K^{+}K^{-}$ decays, where the $f_{0}(980)$ signal in both the $\pi^{+}\pi^{-}$ and $K^{+}K^{-}$ distributions was found. In Ref. \cite{Molina:2019udw}, the decay process $D_{s}^{+} \rightarrow \pi^{+}\pi^{0}\eta$ was studied, and it was found that the $a_{0}(980)$ resonance could be produced via $W$-internal emission, but no need to invoke the $W$-annihilation process, which solved the puzzle of the abnormally large decay rate observed for this decay mode. Continuation of work in this direction is done in Refs. \cite{Duan:2020vye,Ling:2021qzl}. More theoretical studies on three-body decay of $D$ meson with ChUA method can be seen in Refs. \cite{Hsiao:2019ait,Toledo:2020zxj,Roca:2020lyi, Ikeno:2021kzf,Wang:2021nxz}.

In the present work, we study the final state interactions of the singly Cabibbo-suppressed decays of  $D^{0} \rightarrow \pi^{0}\pi^{0}\pi^{0}, \pi^{0}\pi^{0}\eta, \pi^{0}\eta\eta$ with ChUA, where we first get the potential kernel for the hadron-hadron interaction from chiral Lagrangians \cite{Gasser:1984gg}, then solve the Bethe-Salpeter equation in coupled channels. Furthermore, one can calculate branching fractions of decay channels and make predictions for invariant mass spectra. Within the ChUA, the scalar resonances dynamically generate from the hadron-hadron interaction and qualify as molecular states. In our cases, we will consider the contributions of the $f_{0}(500)$, $f_{0}(980)$ and $a_{0}(980)$ states in the two-body final state interactions, and look for their signals in the $\pi^{0}\pi^{0}$ and $\pi^{0}\eta$ components. Furthermore, the experimental results indicated that the decays of  $D^{0} \rightarrow \pi^{0}\pi^{0}\pi^{0}, \pi^{0}\pi^{0}\eta, \pi^{0}\eta\eta$ were dominated by the $W$-internal emission and $W$-exchange mechanism \cite{BESIII:2018hui}. According to the analysis in Refs. \cite{Chau:1982da,Chau:1987tk,Molina:2019udw}, the order of weak decays strength based on topological classification follows as $W$-external emission, $W$-internal emission, $W$-exchange, $W$-annihilation, horizontal $W$-loop and vertical $W$-loop. Since the $W$-external emission has no contribution to these decays, we will only consider the contribution of the $W$-internal emission and omit the $W$-exchange mechanism.

This paper is organized as follows. In Section \ref {sec:Formalism}, we will introduce the formalism of final state interactions under the ChUA. Then, we show the results of the $\pi^{0}\pi^{0}$ and $\pi^{0}\eta$ invariant mass distributions and the ratios of branching fractions for the decays $[D^{0} \rightarrow f_{0}(980) \pi^{0}, f_{0}(980) \rightarrow \pi^{0}\pi^{0}]$, $[D^{0} \rightarrow a_{0}(980) \eta, a_{0}(980) \rightarrow \pi^{0}\eta]$, $[D^{0} \rightarrow a_{0}(980) \pi^{0}, a_{0}(980) \rightarrow \pi^{0}\eta]$, $[D^{0} \rightarrow f_{0}(500) \eta, f_{0}(500) \rightarrow \pi^{0}\pi^{0}]$ and $[D^{0} \rightarrow f_{0}(980) \eta, f_{0}(980) \rightarrow \pi^{0}\pi^{0}]$ in Section \ref {sec:Results}. The conclusion is made in Section \ref {sec:Conclusions}.

\section{Formalism}
\label{sec:Formalism}

\begin{figure}
	\begin{subfigure}{0.45\textwidth}
		\centering
		\includegraphics[width=1\linewidth,trim=150 530 180 120,clip]{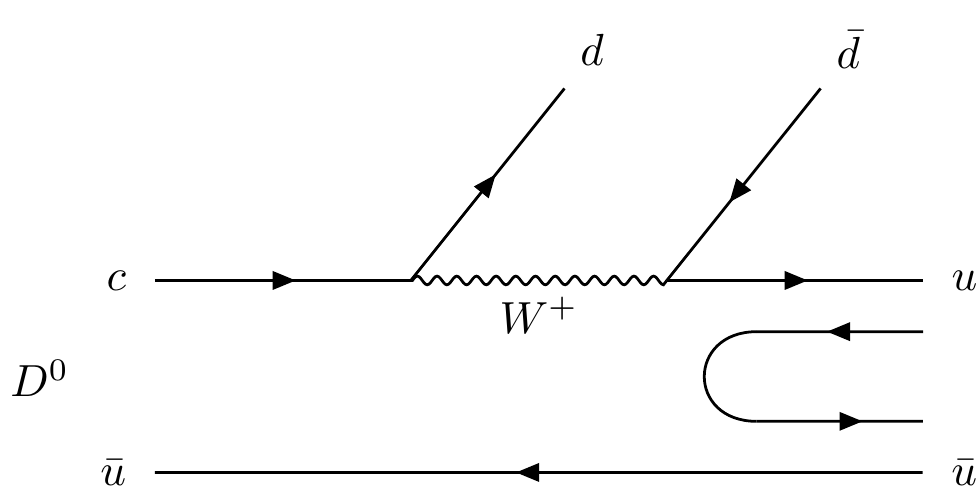} 
		\caption{\footnotesize}
		\label{fig:Feynman1}
	\end{subfigure}
	\quad
	\quad
	\begin{subfigure}{0.45\textwidth}  
		\centering 
		\includegraphics[width=1\linewidth,trim=150 530 180 120,clip]{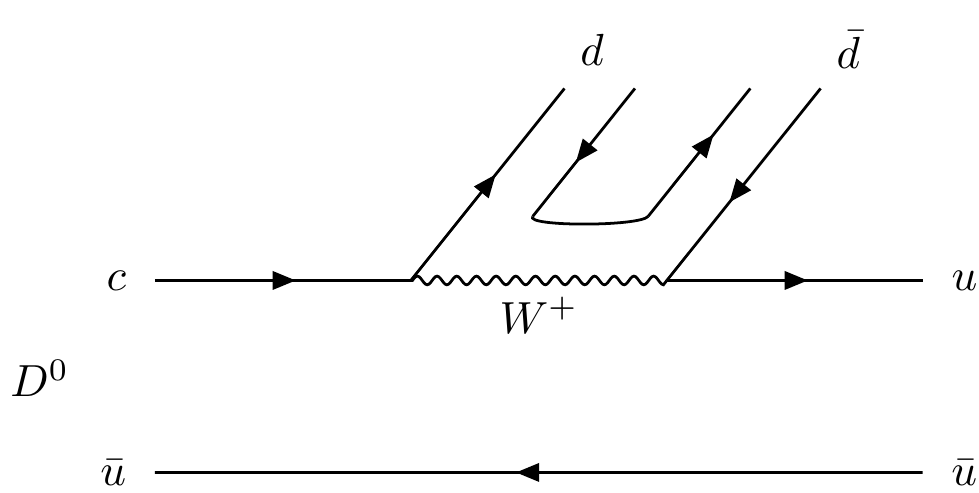} 
		\caption{\footnotesize}
		\label{fig:Feynman2}  
	\end{subfigure}	
	\begin{subfigure}{0.45\textwidth}  
		\centering 
		\includegraphics[width=1\linewidth,trim=150 530 180 120,clip]{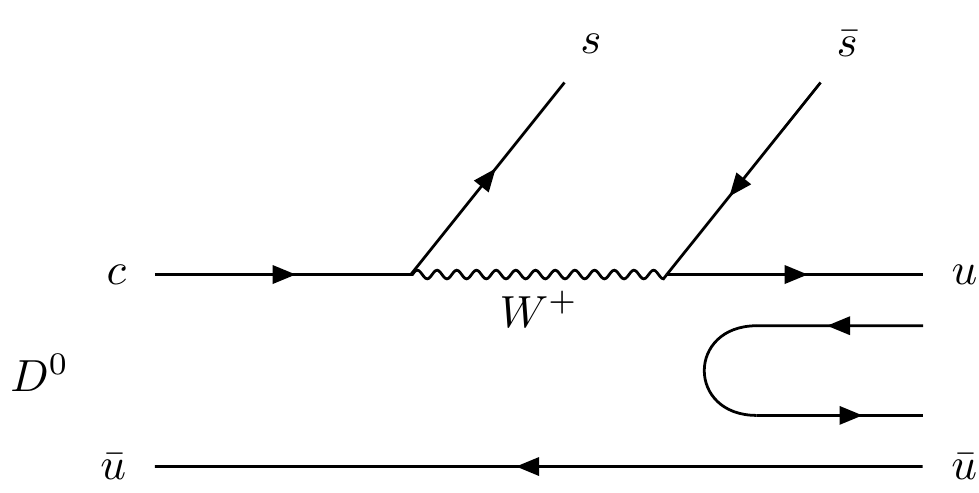} 
		\caption{\footnotesize}
		\label{fig:Feynman3}  
	\end{subfigure}
	\quad
	\quad
	\begin{subfigure}{0.45\textwidth}  
		\centering 
		\includegraphics[width=1\linewidth,trim=150 530 180 120,clip]{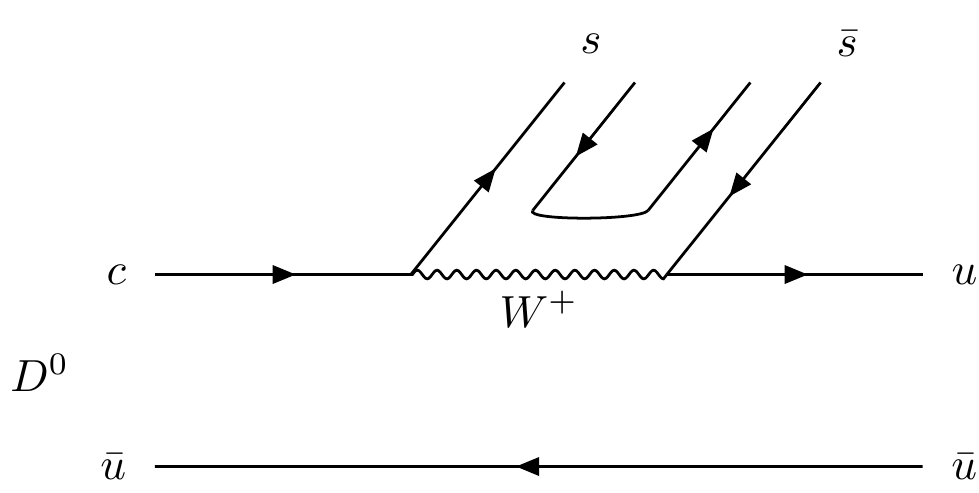} 
		\caption{\footnotesize}
		\label{fig:Feynman4}  
	\end{subfigure}
	\caption{\footnotesize Dominant diagrams for the $D^{0}$ decays with $W$-internal emission. (a) The creations of $d\bar{d}$ and $u\bar{u}$ quarks, then $u\bar{u}$ pair hadronizes into a final pseudoscalar meson pair. (b) The creations of $d\bar{d}$ and $u\bar{u}$ quarks, then $d\bar{d}$ pair hadronizes into a final pseudoscalar meson pair. (c) The creations of $s\bar{s}$ and $u\bar{u}$ quarks, then $u\bar{u}$ pair hadronizes into a final pseudoscalar meson pair. (d) The creations of $s\bar{s}$ and $u\bar{u}$ quarks, then $s\bar{s}$ pair hadronizes into a final pseudoscalar meson pair.}
	\label{fig:Feynman}
\end{figure}

The weak decays $D^{0} \rightarrow \pi^{0}\pi^{0}\pi^{0}, \pi^{0} \pi^{0}\eta, \pi^{0}\eta\eta$ can be described by the Feynman diagrams at quark level by means of $W$-internal emission mechanism as shown in Fig. \ref{fig:Feynman}. We consider all the cases in which the final states contain $\pi^{0}$ or $\eta$ meson, and distinguish these three decay processes for analyzing the amplitudes below. First, let us look at the Fig. \ref{fig:Feynman1}, the $c$ quark in $D^{0}$ meson produces a $d$ quark and a $W^{+}$ boson, while the $\bar{u}$ quark remains as a spectator, then the $W^{+}$ boson goes to $u$ and $\bar{d}$ quarks. The final $d\bar{d}$ pair quarks can form $\pi^{0}$ or $\eta$ meson and the $u\bar{u}$ quarks hadronize by adding an extra $q\bar{q}(\bar{u}u+\bar{d}d+\bar{s}s)$ with the quark pairs created from the vacuum as depicted in Fig. \ref{fig:Feynman1}. This hadronization process can be expressed as

\begin{equation}
	\begin{aligned}
		H^{(a)} & = V_{P} V_{cd} V_{ud} 
		                \left((d \bar{d} \rightarrow \frac{-1}{\sqrt{2}}\pi^{0}) [u\bar{u}\rightarrow u\bar{u} \cdot  (\bar{u}u+\bar{d}d+\bar{s}s)] 
		\right.\\ & \left.+(d \bar{d} \rightarrow \frac{1}{\sqrt{6}}\eta) [u\bar{u}\rightarrow u\bar{u} \cdot  (\bar{u}u+\bar{d}d+\bar{s}s)] \right).
	\end{aligned}
	\label{eq:Ha1}
\end{equation}

Contrarily, the $u\bar{u}$ pair quarks also can merge into $\pi^{0}$ or $\eta$ meson and the $d\bar{d}$ quarks hadronize, which are shown in Fig. \ref{fig:Feynman2}, The hadronization process is formulated as

\begin{equation}
	\begin{aligned}
		H^{(b)} & = V_{P} V_{cd} V_{ud}
		\left( (u \bar{u} \rightarrow \frac{1}{\sqrt{2}}\pi^{0}) [d\bar{d}\rightarrow d\bar{d} \cdot  (\bar{u}u+\bar{d}d+\bar{s}s)] 
		\right.\\ & \left.+(u \bar{u} \rightarrow \frac{1}{\sqrt{6}}\eta) [d\bar{d}\rightarrow d\bar{d} \cdot  (\bar{u}u+\bar{d}d+\bar{s}s)] 
		 \right).
	\end{aligned}
	\label{eq:Hb1}
\end{equation}

In Fig. \ref{fig:Feynman3}, the $\bar u$ quark also remains as a spectator, and the $c$ quark decays into an $s$ quark and a $W^{+}$ boson, then the $W^{+}$ boson goes to a $u$ quark and an $\bar s$ quark. Then the $s\bar{s}$ quark pair can merge into $\eta$ meson. It should be noted that unlike other diagrams, the $s\bar{s}$ quarks cannot form $\pi^{0}$ in this case. The $u\bar{u}$ quark pair hadronizes with the quark pairs produced from the vacuum $q\bar{q}(\bar{u}u+\bar{d}d+\bar{s}s)$, written as

\begin{equation}
	\begin{aligned}
		H^{(c)} & = V_{P} V_{cs} V_{us}(s \bar{s} \rightarrow \frac{-2}{\sqrt{6}}\eta) [u\bar{u}\rightarrow u\bar{u} \cdot  (\bar{u}u+\bar{d}d+\bar{s}s)].
	\end{aligned}
	\label{eq:Hc1}
\end{equation}

Similar to the process of Fig. \ref{fig:Feynman2}, in Fig. \ref{fig:Feynman4}, the $u\bar{u}$ quark pair merges into $\pi^{0}$ or $\eta$ meson and the $s\bar{s}$ quarks hadronize, we can get

\begin{equation}
	\begin{aligned}
		H^{(d)} & = V_{P} V_{cs} V_{us}
		\left( (u \bar{u} \rightarrow \frac{1}{\sqrt{2}}\pi^{0}) [s\bar{s}\rightarrow s\bar{s} \cdot  (\bar{u}u+\bar{d}d+\bar{s}s)] 
		\right.\\ & \left.+(u \bar{u} \rightarrow \frac{1}{\sqrt{6}}\eta) [s\bar{s}\rightarrow s\bar{s} \cdot  (\bar{u}u+\bar{d}d+\bar{s}s)] 
		\right).
	\end{aligned}
	\label{eq:Hd1}
\end{equation}

In Eqs. (\ref {eq:Ha1}-\ref {eq:Hd1}), $V_{P}$ contains all dynamical factors, which is common to all reactions because of the similar production dynamics, and is called the production vertex, we take it as a constant in the calculation \cite{Wang:2020pem,Ahmed:2020qkv}. The $V_{q_{1}q_{2}}$ is the element of the Cabibbo-Kobayashi-Maskawa (CKM) matrix from $q_{1}$ to $q_{2}$ quark. The factors ${1}/{\sqrt{2}}$, ${-1}/{\sqrt{2}}$ of $\pi^{0}$ and ${1}/{\sqrt{6}}$, ${-2}/{\sqrt{6}}$ of $\eta$ in Eqs. (\ref{eq:Ha1}-\ref{eq:Hd1}), are due to the prefactor of the flavor component of the $\pi^{0}$ and $\eta$, which are taken from

\begin{equation}
	\begin{aligned}
		|\pi^{0}\rangle=\frac{1}{\sqrt{2}}|(u \bar u-d \bar d)\rangle, \quad |\eta \rangle=\frac{1}{\sqrt{6}}|(u \bar u+d \bar d-2s\bar s)\rangle.
	\end{aligned}
	\label{eq:pieta}
\end{equation}

Then we define the matrix $M$ for the $q\bar{q}$ elements

\begin{equation}
	\begin{aligned}
		M=\left(\begin{array}{lll}{u \bar{u}} & {u \bar{d}} & {u \bar{s}} \\ {d \bar{u}} & {d \bar{d}} & {d \bar{s}} \\ {s \bar{u}} & {s \bar{d}} & {s \bar{s}}\end{array}\right).
	\end{aligned}
	\label{eq:M}
\end{equation}

\noindent  
So we can easily get the following formulae for the hadronization process

\begin{equation}
	\begin{aligned} 
		u\bar{u} \cdot (\bar{u}u+\bar{d}d+\bar{s}s)=(M \cdot M)_{11},
	\end{aligned} 
	\label{eq:uu1}
\end{equation}

\begin{equation}
	\begin{aligned} 
		d\bar{d} \cdot (\bar{u}u+\bar{d}d+\bar{s}s)=(M \cdot M)_{22},
	\end{aligned} 
	\label{eq:dd1}
\end{equation}

\begin{equation}
	\begin{aligned} 
		s\bar{s} \cdot (\bar{u}u+\bar{d}d+\bar{s}s)=(M \cdot M)_{33}.
	\end{aligned} 
	\label{eq:ss1}
\end{equation}

\noindent
In terms of the pseudoscalar mesons the SU(3) matrix $M$ is given by

\begin{equation}
	\begin{aligned}
		\Phi=\left(\begin{array}{ccc}{\frac{1}{\sqrt{2}} \pi^{0}+\frac{1}{\sqrt{6}} \eta} & {\pi^{+}} & {K^{+}} \\ {\pi^{-}} & {-\frac{1}{\sqrt{2}} \pi^{0}+\frac{1}{\sqrt{6}} \eta} & {K^{0}} \\ {K^{-}} & {\bar{K}^{0}} & {-\frac{2}{\sqrt{6}} \eta}\end{array}\right),
	\end{aligned}
	\label{eq:Phi}
\end{equation} 

\noindent  
where we take $\eta \equiv \eta_{8}$,  and the singlet of SU(3) components $\eta_{1}$ is removed since it does not lead to any interaction in chiral perturbation theory \cite{Liang:2014tia}. The hadronization process in quark level in Eqs. (\ref{eq:uu1}-\ref{eq:ss1}) can be accomplished to the hadron level in terms of two pseudoscalar mesons, given by

\begin{equation}
	\begin{aligned}
		(M \cdot M)_{11}= (\Phi \cdot \Phi)_{11} = \pi^{+}\pi^{-} + \frac{1}{2}\pi^{0}\pi^{0} + \frac{1}{\sqrt{3}}\pi^{0}\eta + K^{+}{K}^{-}+\frac{1}{6} \eta \eta, 
	\end{aligned}
	\label{eq:uu2}
\end{equation} 

\begin{equation}
	\begin{aligned}
		(M \cdot M)_{22}= (\Phi \cdot \Phi)_{22} = \pi^{+}\pi^{-} + \frac{1}{2}\pi^{0}\pi^{0} - \frac{1}{\sqrt{3}}\pi^{0}\eta + K^{0}\bar{K}^{0}+\frac{1}{6} \eta \eta, 
	\end{aligned}
	\label{eq:dd2}
\end{equation} 

\begin{equation}
	\begin{aligned}
		(M \cdot M)_{33}= (\Phi \cdot \Phi)_{33}=K^{+}{K}^{-}+ K^{0}\bar{K}^{0}+\frac{2}{3} \eta \eta.
	\end{aligned}
	\label{eq:ss2}
\end{equation} 

Then, after the hadronization, we get the final states with $\pi^{0}$ or $\eta$ as follows

\begin{equation}
	\begin{aligned}
		H^{(a)} & = V_{P} V_{cd} V_{ud} 
		\left((\frac{-1}{\sqrt{2}}\pi^{0}) (\pi^{+}\pi^{-} + \frac{1}{2}\pi^{0}\pi^{0} + \frac{1}{\sqrt{3}}\pi^{0}\eta + K^{+}{K}^{-}+\frac{1}{6} \eta \eta) 
		\right.\\ & \left.+\frac{1}{\sqrt{6}}\eta(\pi^{+}\pi^{-} + \frac{1}{2}\pi^{0}\pi^{0} + \frac{1}{\sqrt{3}}\pi^{0}\eta + K^{+}{K}^{-}+\frac{1}{6} \eta \eta)
		\right),
	\end{aligned}
	\label{eq:Ha2}
\end{equation}

\begin{equation}
	\begin{aligned}
		H^{(b)} & = V_{P} V_{cd} V_{ud}
		\left( \frac{1}{\sqrt{2}}\pi^{0} (\pi^{+}\pi^{-} + \frac{1}{2}\pi^{0}\pi^{0} - \frac{1}{\sqrt{3}}\pi^{0}\eta + K^{0}\bar{K}^{0}+\frac{1}{6} \eta \eta)
		\right.\\ & \left.+ \frac{1}{\sqrt{6}}\eta (\pi^{+}\pi^{-} + \frac{1}{2}\pi^{0}\pi^{0} - \frac{1}{\sqrt{3}}\pi^{0}\eta + K^{0}\bar{K}^{0}+\frac{1}{6} \eta \eta)  
		\right),
	\end{aligned}
	\label{eq:Hb2}
\end{equation}

\begin{equation}
	\begin{aligned}
		H^{(c)} & = V_{P} V_{cs} V_{us}(\frac{-2}{\sqrt{6}}\eta) (\pi^{+}\pi^{-} + \frac{1}{2}\pi^{0}\pi^{0} + \frac{1}{\sqrt{3}}\pi^{0}\eta + K^{+}{K}^{-}+\frac{1}{6} \eta \eta),
	\end{aligned}
	\label{eq:Hc2}
\end{equation}

\begin{equation}
	\begin{aligned}
		H^{(d)} & = V_{P} V_{cs} V_{us}
		\left( \frac{1}{\sqrt{2}}\pi^{0} (K^{+}{K}^{-}+ K^{0}\bar{K}^{0}+\frac{2}{3} \eta \eta)
		\right.\\ & \left.+\frac{1}{\sqrt{6}}\eta (K^{+}{K}^{-}+ K^{0}\bar{K}^{0}+\frac{2}{3} \eta \eta) 
		\right).
	\end{aligned}
	\label{eq:Hd2}
\end{equation}

\begin{figure}
	\begin{subfigure}{0.45\textwidth}
		\centering
		\includegraphics[width=1\linewidth,trim=150 530 180 120,clip]{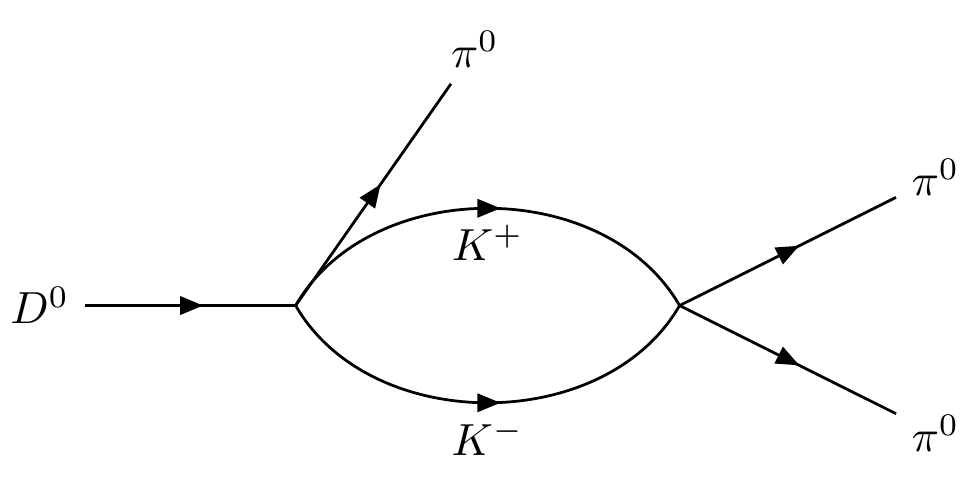} 
		\caption{\footnotesize}
		\label{fig:Scatter11}
	\end{subfigure}
	\quad
	\quad
	\begin{subfigure}{0.45\textwidth}  
		\centering 
		\includegraphics[width=1\linewidth,trim=150 530 180 120,clip]{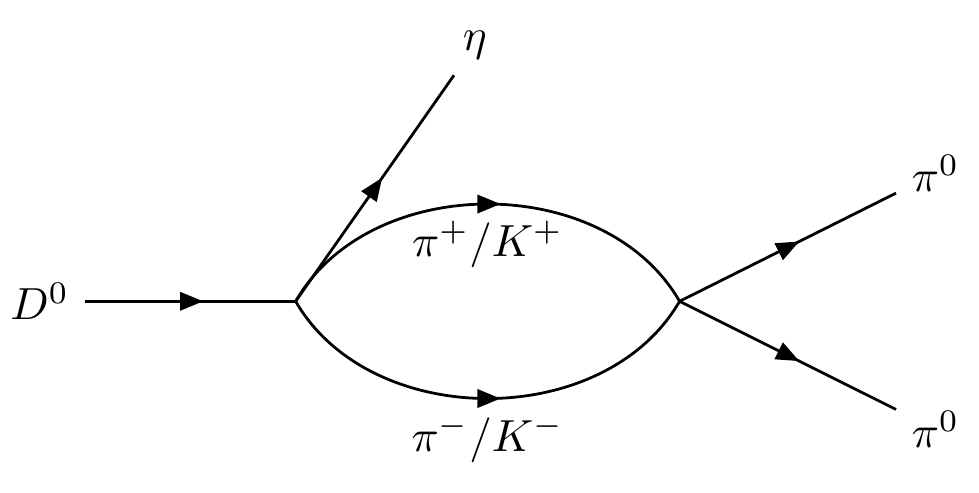} 
		\caption{\footnotesize}
		\label{fig:Scatter21}  
	\end{subfigure}	
	\quad
	\quad
	\begin{subfigure}{0.45\textwidth}  
		\centering 
		\includegraphics[width=1\linewidth,trim=150 530 180 120,clip]{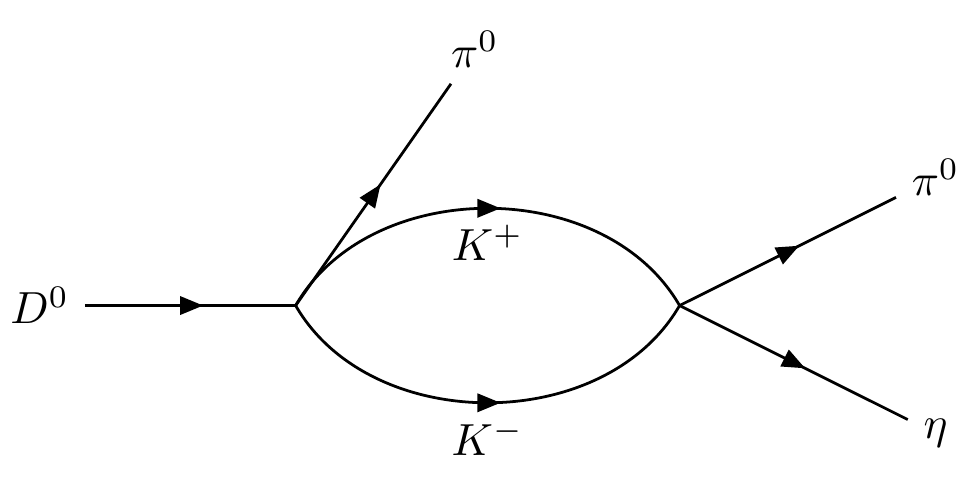} 
		\caption{\footnotesize}
		\label{fig:Scatter22}  
	\end{subfigure}	
	\quad
	\quad
	\begin{subfigure}{0.45\textwidth}  
		\centering 
		\includegraphics[width=1\linewidth,trim=150 530 180 120,clip]{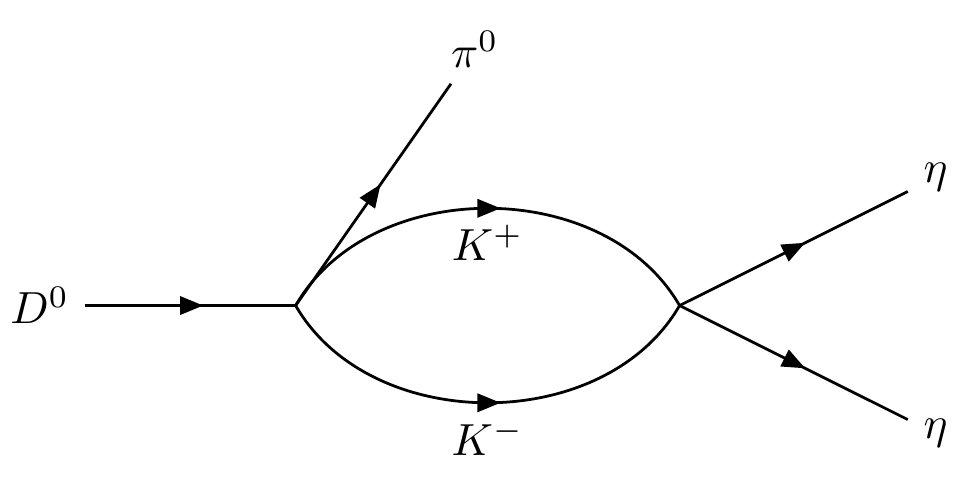} 
		\caption{\footnotesize}
		\label{fig:Scatter31}  
	\end{subfigure}	
	\quad
	\quad
	\begin{subfigure}{1\textwidth}  
		\centering 
		\includegraphics[width=0.45\linewidth,trim=150 530 180 120,clip]{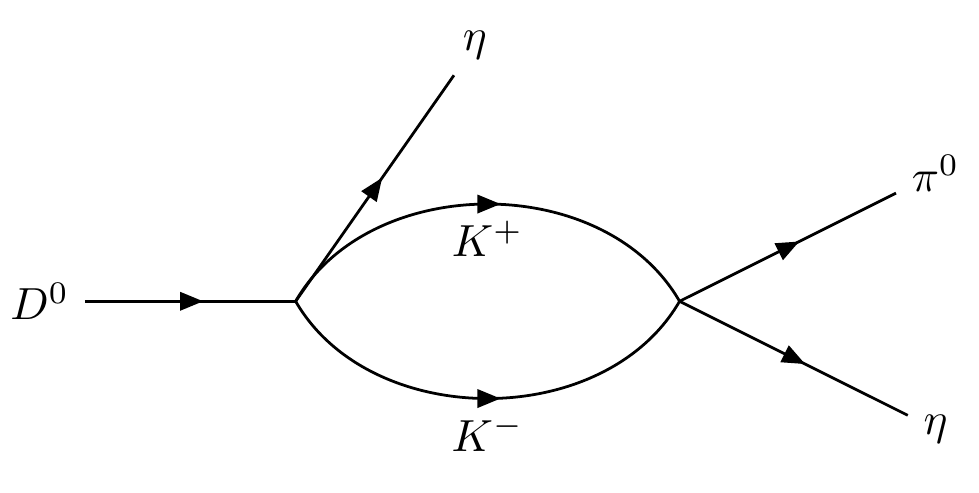} 
		\caption{\footnotesize}
		\label{fig:Scatter32}  
	\end{subfigure}		
	\caption{\footnotesize Diagrammatic representation of the mechanisms for the final state interactions of the meson pairs. (a) The final state interaction of the $D^{0} \rightarrow \pi^{0}\pi^{0}\pi^{0}$ decay, (b) and (c) the final state interaction of the $D^{0} \rightarrow \pi^{0} \pi^{0}\eta$ decay, (d) and (e) the final state interaction of the $D^{0} \rightarrow \pi^{0}\eta\eta$ decay.}
	\label{fig:Scatter}
\end{figure}

Note that the elements of the CKM matrix are $V_{cd}=-V_{us}$, $V_{ud}=V_{cs}$ \cite{Zyla:2020pdg}, leading to $V_{cd}V_{ud}=-V_{us}V_{cs}$. Thus we get the total contributions for Figs. \ref{fig:Feynman1}-\ref{fig:Feynman4},

\begin{equation}
	\begin{aligned}
		H & = H^{(a)}+H^{(b)}+H^{(c)}+H^{(d)}
		\\ & =  C \left( -\sqrt{2}\pi^{0}K^{+}K^{-}+\frac{4}{\sqrt{6}}\pi^{+}\pi^{-}\eta +\frac{2}{\sqrt{6}}\eta K^{+}K^{-} \right),
	\end{aligned}
	\label{eq:H}
\end{equation}

\noindent
where $C$ is a global factor, which absorbs the production vertex $V_{P}$ and the elements of the CKM matrix $V_{cd} V_{ud}$ or $V_{cs} V_{us}$, and also includes the normalization factor used to match the events of the experimental data. Note that, there are no final states $\pi^{0}\pi^{0}\pi^{0}$, $\pi^{0}\pi^{0}\eta$ or $\pi^{0}\eta\eta$ directly produced that we want in the $D^{0}$ decay, since these final states are cancelled by each other in the summation of Eqs. (\ref{eq:Ha2}-\ref{eq:Hd2}). However, upon rescattering of the terms in Eq. (\ref{eq:H}) we can get them via the final state interaction, as depicted in Fig. \ref{fig:Scatter}. Then we get the amplitudes for the $D^{0} \rightarrow \pi^{0}\pi^{0}\pi^{0}$ decay 

\begin{equation}
	\begin{aligned} 
		t_{D^{0} \rightarrow \pi^{0}\pi^{0}\pi^{0}}=-\sqrt{2} C G_{K^{+}K^{-}}(M_{inv}(\pi^{0} \pi^{0})) T_{K^{+}K^{-} \rightarrow \pi^{0}\pi^{0}}(M_{inv}(\pi^{0}\pi^{0})),
	\end{aligned}
	\label{eq:amplitudes1}
\end{equation}

\noindent
for the $D^{0} \rightarrow \pi^{0} \pi^{0}\eta$ decay

\begin{equation}
	\begin{aligned} 
		t_{D^{0} \rightarrow \pi^{0}\pi^{0}\eta}=C & \left [ \frac{4}{\sqrt{6}}G_{\pi^{+} \pi^{-}}(M_{inv}(\pi^{0} \pi^{0})) T_{\pi^{+} \pi^{-} \rightarrow \pi^{0} \pi^{0}}(M_{inv}(\pi^{0} \pi^{0}))
		\right.\\ & \left.+\frac{2}{\sqrt{6}}G_{K^{+}K^{-}}(M_{inv}(\pi^{0} \pi^{0}))  T_{K^{+}K^{-} \rightarrow \pi^{0} \pi^{0}}(M_{inv}(\pi^{0} \pi^{0})) 
		\right.\\ & \left. -\sqrt{2} G_{K^{+}K^{-}}(M_{inv}(\pi^{0} \eta)) T_{K^{+}K^{-} \rightarrow \pi^{0} \eta}(M_{inv}(\pi^{0} \eta))
		\right],
	\end{aligned}
	\label{eq:amplitudes2}
\end{equation}

\noindent
and the one for the $D^{0} \rightarrow \pi^{0}\eta\eta$ decay

\begin{equation}
	\begin{aligned} 
		t_{D^{0} \rightarrow \pi^{0}\eta\eta}=C & \left [ -\sqrt{2}G_{K^{+}K^{-}}(M_{inv}(\eta\eta))  T_{K^{+}K^{-} \rightarrow \eta\eta}(M_{inv}(\eta\eta))
		\right.\\ & \left. +\frac{2}{\sqrt{6}} G_{K^{+}K^{-}}(M_{inv}(\pi^{0} \eta)) T_{K^{+}K^{-} \rightarrow \pi^{0} \eta}(M_{inv}(\pi^{0} \eta))
		\right].
	\end{aligned}
	\label{eq:amplitudes3}
\end{equation}

For isospin $I = 0$, we consider five coupled channels, $\pi^{+}\pi^{-}$ (1), $\pi^{0}\pi^{0}$ (2), $K^{+}K^{-}$ (3), $K^{0}\bar {K}^{0}$ (4) and $\eta\eta$ (5). For isospin $I = 1$, we consider the contribution of three coupled channels, $K^{+}K^{-}$ (1), $K^{0}\bar{K}^{0}$ (2) and $\pi^{0}\eta$ (3). Thus, the $\pi^{+}\pi^{-}$, $\pi^{0}\pi^{0}$ and $\eta\eta$ channels only contribute to $I = 0$ \cite{Roca:2020lyi}, and the $\pi^{0}\eta$ channel only contributes to $I = 1$. The $K^{+}K^{-}$ and $K^{0}\bar{K}^{0}$ channels contribute to both isospins $I = 0$ and $I = 1$, taking into account the isospin decomposition of the $K\bar K$ states

\begin{equation}
	\begin{aligned} 
		\left|K^{+} K^{-}\right\rangle=-\frac{1}{\sqrt{2}}|K \bar{K}\rangle_{I=1, I_{3}=0}-\frac{1}{\sqrt{2}}|K \bar{K}\rangle_{I=0, I_{3}=0},
	\end{aligned}
	\label{eq:KK1}
\end{equation}

\begin{equation}
	\begin{aligned} 
		\left|K^{0} \bar K^{0}\right\rangle=\frac{1}{\sqrt{2}}|K \bar{K}\rangle_{I=1, I_{3}=0}-\frac{1}{\sqrt{2}}|K \bar{K}\rangle_{I=0, I_{3}=0},
	\end{aligned}
	\label{eq:KK2}
\end{equation}

\noindent
where we have used the convention that $\left|K^{+}\rangle=-\right| 1/2,\,1/2\rangle$ for the isospin basis \cite{Oller:1997ti}. Thus, the final state interaction amplitudes in Eqs. (\ref{eq:amplitudes1}-\ref{eq:amplitudes3}) are give by

\begin{equation}
	\begin{aligned} 
		T_{K^{+}K^{-}\rightarrow \pi^{0} \pi^{0}}=\frac{1}{2} \left(T_{K^{0} \bar K^{0} \rightarrow \pi^{0} \pi^{0}}+T_{K^{+}{K}^{-} \rightarrow \pi^{0} \pi^{0}}\right),
	\end{aligned}
	\label{eq:KKPIPI}
\end{equation}

\begin{equation}
	\begin{aligned} 
		T_{K^{+}K^{-}\rightarrow \eta \eta}=\frac{1}{2} \left(T_{K^{0} \bar K^{0} \rightarrow \eta \eta}+T_{K^{+}{K}^{-} \rightarrow \eta \eta}\right),
	\end{aligned}
	\label{eq:KKETAETA}
\end{equation}

\begin{equation}
	\begin{aligned} 
		T_{K^{+}K^{-}\rightarrow \pi^{0} \eta}=\frac{1}{2} \left(T_{K^{+}{K}^{-} \rightarrow \pi^{0} \eta}-T_{K^{0} \bar K^{0} \rightarrow \pi^{0} \eta}\right).
	\end{aligned}
	\label{eq:KKPIETA}
\end{equation}

In addition, the $G_ {kk}$ in Eqs. (\ref{eq:amplitudes1}-\ref{eq:amplitudes3}) is the loop functions of two meson propagators, which is given by 

\begin{equation}
	\begin{aligned}
		G _ {kk} ( s ) = i \int \frac { d ^ { 4 } q } { ( 2 \pi ) ^ { 4 } } \frac { 1 } { q ^ { 2 } - m _ { 1 } ^ { 2 } + i \varepsilon } \frac { 1 } { \left( p _ { 1 } + p _ { 2 } - q \right) ^ { 2 } - m _ { 2 } ^ { 2 } + i \varepsilon }  \text{ ,}
	\end{aligned}
	\label{eq:propagators}
\end{equation}

\noindent 
where $p_{1}$ and $p_{2}$ are the four-momenta of the two initial particles, respectively, and $m_{1}$ and $m_{2}$ are the masses of the two intermediate particles. The integral of this equation is logarithmically divergent, and we take the formula of dimensional regularization method to solve this singular integral \cite{Oller:2000ma,Gamermann:2006nm,Alvarez-Ruso:2010rqm,Guo:2016zep}

\begin{equation}
	\begin{aligned}
		G_{kk}(s)=& \frac{1}{16 \pi^{2}}\left\{a_{k}(\mu)+\ln \frac{m_{1}^{2}}{\mu^{2}}+\frac{m_{2}^{2}-m_{1}^{2}+s}{2 s} \ln \frac{m_{2}^{2}}{m_{1}^{2}}\right.\\
		&+\frac{q_{cmk}(s)}{\sqrt{s}}\left[\ln \left(s-\left(m_{2}^{2}-m_{1}^{2}\right)+2 q_{cmk}(s) \sqrt{s}\right)\right.\\
		&+\ln \left(s+\left(m_{2}^{2}-m_{1}^{2}\right)+2 q_{cmk}(s) \sqrt{s}\right) \\
		&-\ln \left(-s-\left(m_{2}^{2}-m_{1}^{2}\right)+2 q_{cmk}(s) \sqrt{s}\right) \\
		&\left.\left.-\ln \left(-s+\left(m_{2}^{2}-m_{1}^{2}\right)+2 q_{cmk}(s) \sqrt{s}\right)\right]\right\},
	\end{aligned}
	\label{eq:DR}
\end{equation}

\noindent
where $\mu$ is the regularization scale, $a_{k}(\mu)$ the subtraction constant, and we take $\mu=0.6$ GeV as in Ref. \cite{Duan:2020vye}. As discussed in Ref. \cite{Duan:2020vye}, following the Eq. (17) of Ref. \cite{Oller:2000fj}, one has

\begin{equation}
	\begin{aligned}
		a_{k}(\mu)=-2 \log \left(1+\sqrt{1+\frac{m_{k}^{2}}{\mu^{2}}}\right)+\cdots,
	\end{aligned}
\end{equation}

\noindent
where index $k$ represents the coupled channels, $m_{k}$ is the mass of larger-mass meson in the coupled channels \footnote{Note that in Ref. \cite{Oller:2000fj} $m_{k}$ is the mass of baryon in meson-baryon coupled channels.}. And then we get the values of the subtraction constants $a_{\pi^{+}\pi^{-}}=-1.41$, $a_{\pi^{0}\pi^{0}}=-1.41$, $a_{K^{+}K^{-}}=-1.66$, $a_{K^{0}\bar{K}^{0}}=-1.66$, $a_{\eta\eta}=-1.71$ and $a_{\pi^{0}\eta}=-1.71$. Besides, $q_{cmk}(s)$ is the three momentum of the particle in the center-of-mass frame, given by 

\begin{equation}
	\begin{aligned}
		q_{cmk}(s)=\frac{\lambda^{1 / 2}\left(s, m_{1}^{2}, m_{2}^{2}\right)}{2 \sqrt{s}},
	\end{aligned}
	\label{eq:qcmk}
\end{equation}

\noindent
with the usual Källen triangle function $\lambda(a, b, c)=a^{2}+b^{2}+c^{2}-2(a b+a c+b c)$.

Besides, $T_{ij}$ is element of the scattering amplitude matrices for the transitions of channel $i \rightarrow j$ in the ChUA evaluated by the coupled channel Bethe-Salpeter equation

\begin{equation}
	\begin{aligned} 
		T = [1-VG]^{-1}V, 
	\end{aligned}
	\label{eq:BSE}
\end{equation}

\noindent
where the matrix $V$ is constructed by the scattering potentials of each coupled channel and obtained from the lowest order chiral Lagrangians. For $I=0$ sector, it is a $5\times 5$ symmetric matrix, which is given by \cite{Liang:2014tia},

\begin{equation}
	\begin{aligned}
		&V_{11}=-\frac{1}{2 f^{2}} s, \quad V_{12}=-\frac{1}{\sqrt{2} f^{2}}\left(s-m_{\pi}^{2}\right), \quad V_{13}=-\frac{1}{4 f^{2}} s ,\\
		&V_{14}=-\frac{1}{4 f^{2}} s, \quad V_{15}=-\frac{1}{3 \sqrt{2} f^{2}} m_{\pi}^{2}, \quad V_{22}=-\frac{1}{2 f^{2}} m_{\pi}^{2} ,\\
		&V_{23}=-\frac{1}{4 \sqrt{2} f^{2}} s, \quad V_{24}=-\frac{1}{4 \sqrt{2} f^{2}} s, \quad V_{25}=-\frac{1}{6 f^{2}} m_{\pi}^{2} ,\\
		&V_{33}=-\frac{1}{2 f^{2}} s, \quad V_{34}=-\frac{1}{4 f^{2}} s ,\\
		&V_{35}=-\frac{1}{12 \sqrt{2} f^{2}}\left(9 s-6 m_{\eta}^{2}-2 m_{\pi}^{2}\right), \quad V_{44}=-\frac{1}{2 f^{2}} s ,\\
		&V_{45}=-\frac{1}{12 \sqrt{2} f^{2}}\left(9 s-6 m_{\eta}^{2}-2 m_{\pi}^{2}\right) ,\\
		&V_{55}=-\frac{1}{18 f^{2}}\left(16 m_{K}^{2}-7 m_{\pi}^{2}\right),
	\end{aligned}
\end{equation}

\noindent
and the one for $I=1$ sector is a $3\times 3$ symmetric matrix \cite{Xie:2014tma},

\begin{equation}
	\begin{aligned}
		&V_{11}=-\frac{1}{2 f^{2}} s, \quad V_{12}=-\frac{1}{4 f^{2}} s , \\
		&V_{13}=-\frac{\sqrt{3}}{12 f^{2}}\left(3s-\frac {8}{3}m_{K}^{2}-\frac {1}{3}m_{\pi}^{2}-m_{\eta}^{2}\right), \quad V_{22}=-\frac{1}{2 f^{2}} s , \\
		&V_{23}=\frac{\sqrt{3}}{12 f^{2}}\left(3s-\frac {8}{3}m_{K}^{2}-\frac {1}{3}m_{\pi}^{2}-m_{\eta}^{2}\right), \quad V_{33}=-\frac{1}{3 f^{2}}m_{\pi}^{2},
	\end{aligned}
\end{equation}

\noindent
where $f$ is the pion decay constant, and we take $f=0.093$ GeV \cite{Oller:1997ti}.

Finally, the formula of double differential width for three-body decay process is given by \cite{Zyla:2020pdg,Roca:2020lyi}

\begin{equation}
	\begin{aligned}
		\frac{d \Gamma}{d M_{12}d M_{23}}=\frac{1}{(2 \pi)^{3}} \frac{M_{12}M_{23}}{8 m_{D^{0}}^{3}}\frac{1}{N} \left|t\right|^{2},
	\end{aligned}
	\label{eq:dGamma}
\end{equation}

\noindent
where $N$ is the number of identical particles in the final states \cite{Roca:2020lyi}, like $N=3$ for the decay $D^{0} \rightarrow \pi^{0}\pi^{0}\pi^{0}$. For the decays $D^{0} \rightarrow \pi^{0}\pi^{0}\eta$ and $D^{0} \rightarrow \pi^{0}\eta\eta$, there are two cases, $N=2$ for the final states of $\pi^{0}\eta$ components in these two decays, $N=1$ for the final states of $\pi^{0}\pi^{0}$ components in the $D^{0} \rightarrow \pi^{0}\pi^{0}\eta$ decay and $\eta\eta$ components in the $D^{0} \rightarrow \pi^{0}\eta\eta$ decay. And $t=t(M_{12})+t(M_{13})+t(M_{23})$, which depends on the invariant masses of two components $M_{12}, M_{13}, M_{23}$, where the indices $1$ to $3$ denote the final meson state accordingly, although only two of these variables are independent since 

\begin{equation}
	\begin{aligned}
		M_{12}^{2}+M_{23}^{2}+M_{13}^{2} = m_{D^{0}}^{2}+m_{1}^{2}+m_{2}^{2}+m_{3}^{2}.
	\end{aligned}
	\label{eq:energyconservation}
\end{equation}

Then the $d \Gamma/dM_{12}$ and $d \Gamma/dM_{23}$ can be obtained by integrating Eq. (\ref{eq:dGamma}) over each of the invariant mass variables. Furthermore, one can obtain $d \Gamma/dM_{13}$ through the Eq. (\ref{eq:energyconservation}).

\section{Results}
\label{sec:Results}

\begin{figure}[htbp]
	\centering
	\includegraphics[width=0.6\linewidth]{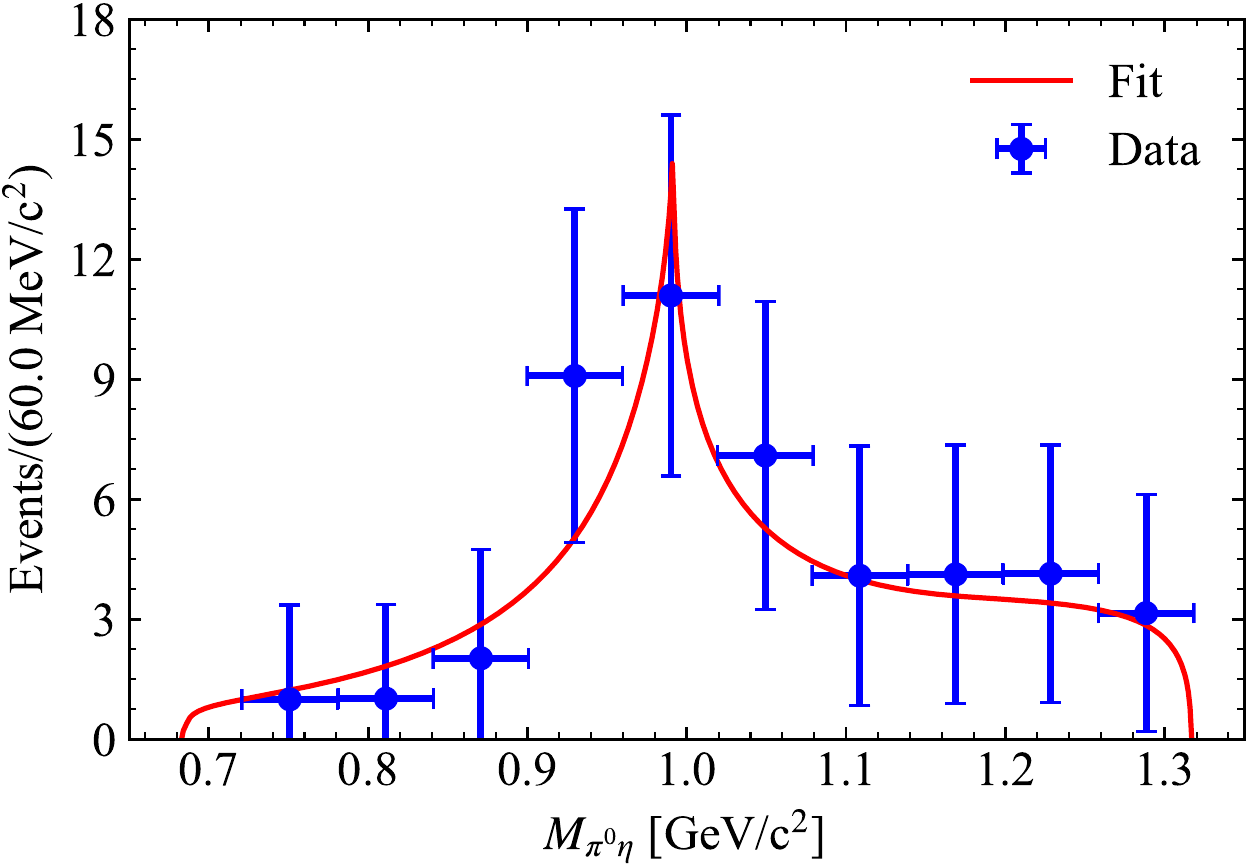} 
	\caption{\footnotesize The $\pi^{0}\eta$ invariant mass distribution of the $D^{0} \rightarrow \pi^{0}\eta\eta$ decay. Parameter $C=918.33$ with the reduced chi-square $\chi^{2}/dof.=1.82/(10-1)=0.20$. Data is taken from \cite{BESIII:2018hui}.}
	\label{fig:3dGamma12}
\end{figure}

In our model, we have only one parameter $C$ for the global normalization in Eqs. (\ref {eq:amplitudes1}-\ref {eq:amplitudes3}). In the introduction, we have mentioned that the BESIII Collaboration had reported the decay of $D^{0} \rightarrow \pi^{0}\eta\eta$,  where the invariant mass spectrum of the $\pi^{0}\eta$ was given in Ref. \cite{BESIII:2018hui}. We first fit the invariant mass spectrum to determine the value of parameter $C$. It is worth emphasizing that the parameter $C$ only determines the overall strength, but does not affect the trend of the curve. Our result is shown in Fig. \ref{fig:3dGamma12}, which clearly shows that the fitting result is in a good agreement with the experimental data. We do not add the resonant state $a_{0}(980)$ to the theoretical formula by hand. The structure of the peak near the threshold of $K\bar K$ in Fig. \ref{fig:3dGamma12} is dynamically generated with the ChUA, where the $a_{0}(980)$ state can be well reproduced. It should be noted that the result shows the typical cusp effect for the $a_{0}(980)$, which is consistent with many calculations \cite{Liang:2016hmr,Debastiani:2016ayp,Molina:2019udw,Ahmed:2020kmp,Toledo:2020zxj}. And in high precision experimental measurements,  this kind of cusp structure appears quite evidently for the $a_{0}(980)$ resonance \cite{Rubin:2004cq,Kornicer:2016axs,Kornicer:2016ywv,BaBar:2021fkz}. The similar situation is also visible in recent lattice QCD simulation \cite{Molina:2019udw}. The experimental data of the $D^{0} \rightarrow \pi^{0}\eta\eta$ decay from BESIII Collaboration did not show this characteristic clearly, because the sampling intervals and errors are very large. It is expected that future experiments can give more accurate measurements. 

\begin{figure}[htbp]
	\centering
	\includegraphics[width=0.6\linewidth]{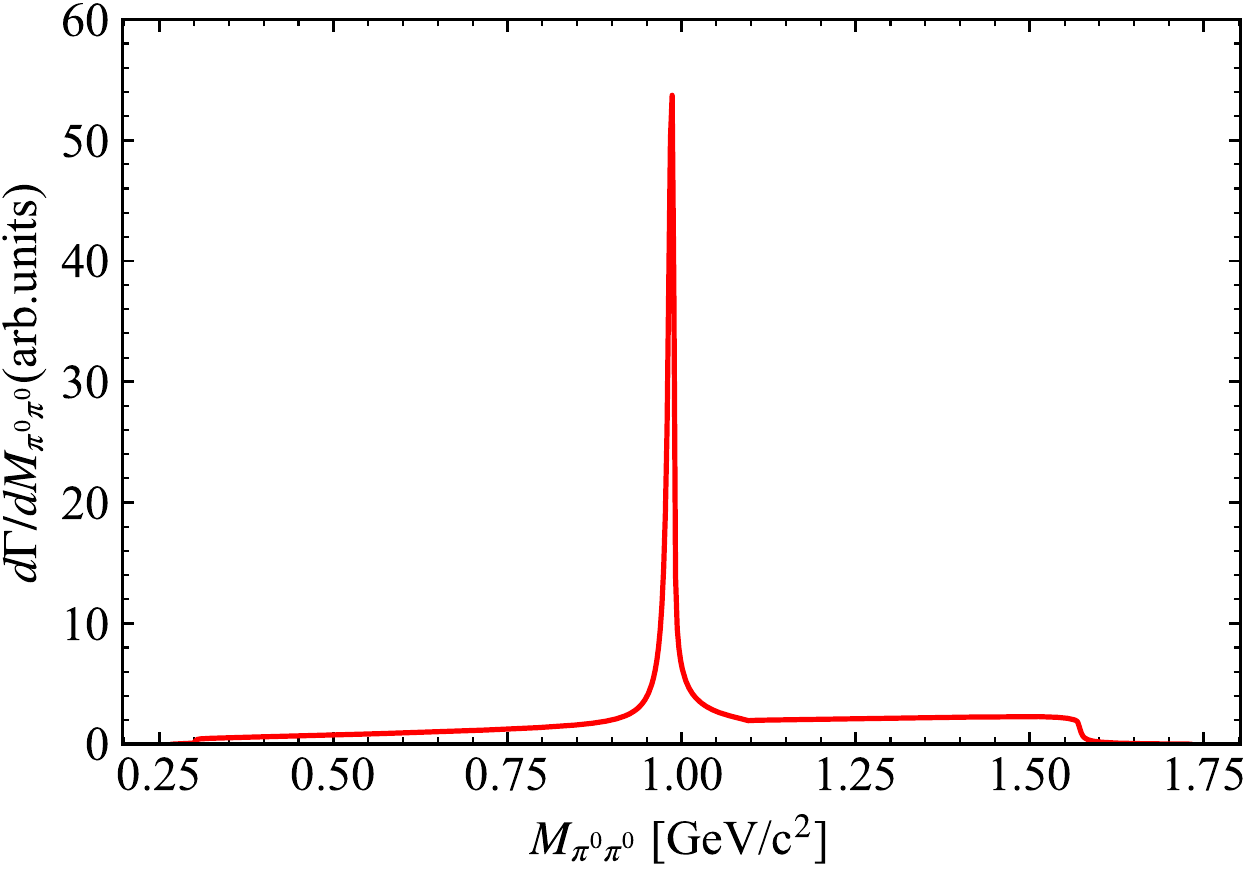} 
	\caption{\footnotesize The $\pi^{0}\pi^{0}$ invariant mass distribution of the $D^{0} \rightarrow \pi^{0}\pi^{0}\pi^{0}$ decay.}
	\label{fig:1dGamma12}
\end{figure}

We show the $\pi^{0}\pi^{0}$ invariant mass distribution of the decay $D^{0} \rightarrow \pi^{0}\pi^{0}\pi^{0}$ in Fig. \ref{fig:1dGamma12}, where the peak of $f_{0}(980)$ rises near the $K\bar K$ threshold with no signal of $f_{0}(500)$ in the invariant mass spectrum. From the results in the ChUA \cite{Oller:1997ti,Ahmed:2020kmp}, we know that the resonance $f_{0}(980)$ is the bound state of $K\bar K$ components, and the $f_{0}(500)$ state is mainly contributed by the $\pi\pi$ channel. The $\pi^{0}\pi^{0}$ invariant mass spectrum of the decay $D^{0} \rightarrow \pi^{0}\pi^{0}\pi^{0}$ is contributed by the amplitude of $K^{+}K^{-} \rightarrow\pi^{0}\pi^{0}$ as shown in Eq. (\ref{eq:amplitudes1}). Thus, the absence of the $f_{0}(500)$ in this result is not surprising, and indicates a different nature of these two states.

\begin{figure}[htbp]
	\begin{subfigure}{0.45\textwidth}
		\centering
		\includegraphics[width=1\linewidth]{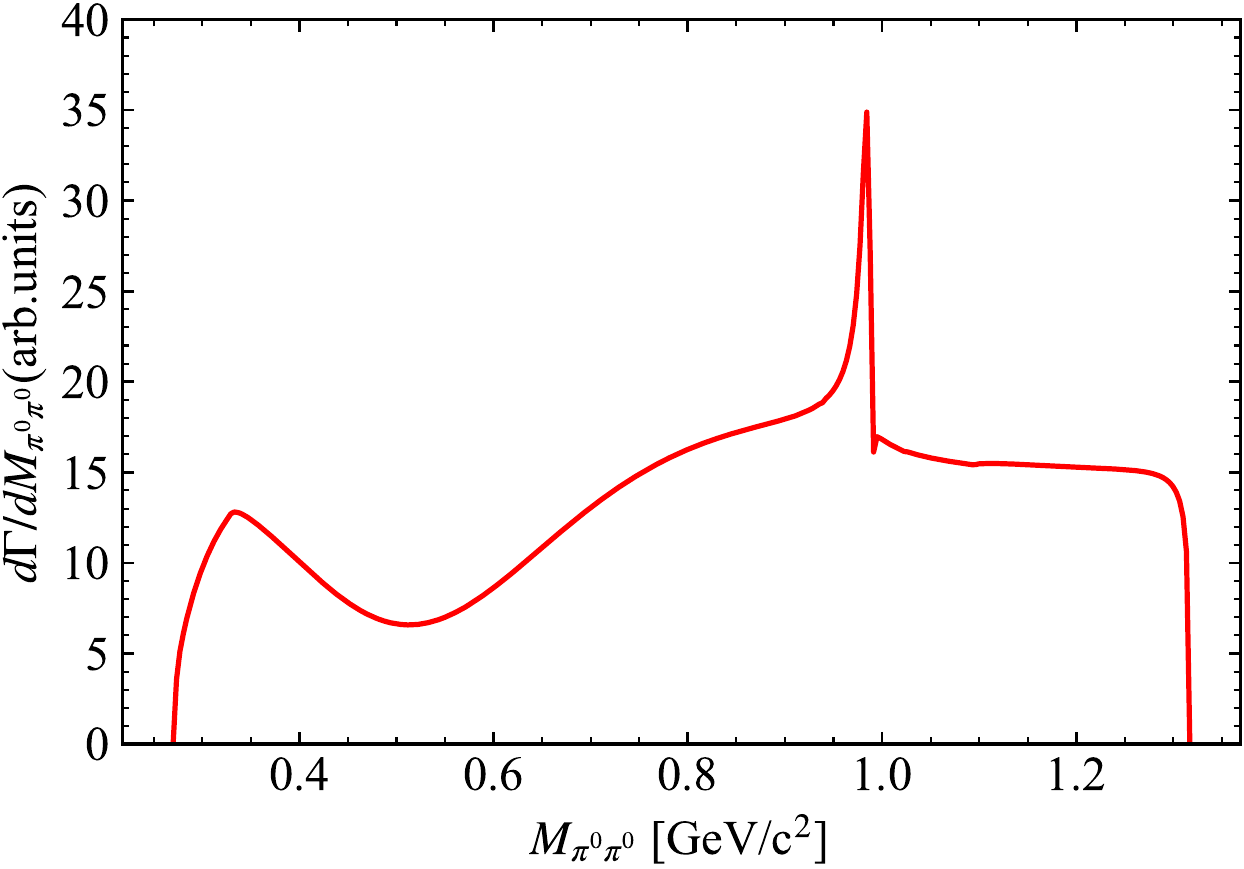} 
		\caption{\footnotesize}
		\label{fig:2dGamma12}
	\end{subfigure}
	\quad
	\quad
	\begin{subfigure}{0.45\textwidth}  
		\centering 
		\includegraphics[width=1\linewidth]{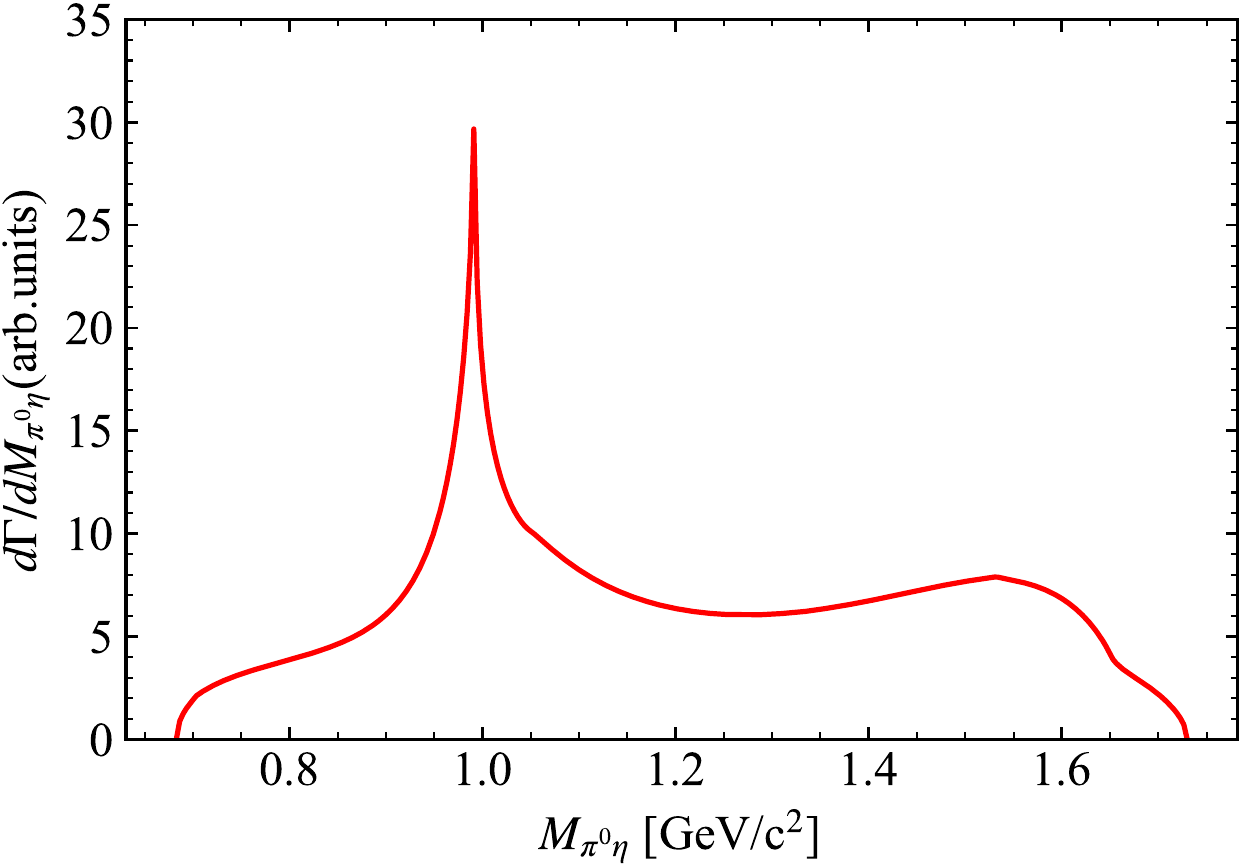} 
		\caption{\footnotesize}
		\label{fig:2dGamma13}  
	\end{subfigure}	
	\caption{\footnotesize The $\pi^{0}\pi^{0}$ (a) and $\pi^{0}\eta$ (b) invariant mass distributions of the $D^{0} \rightarrow \pi^{0}\pi^{0}\eta$ decay.}
	\label{fig:2dGamma}
\end{figure}

\begin{figure}[htbp]
	\centering
	\includegraphics[width=0.6\linewidth]{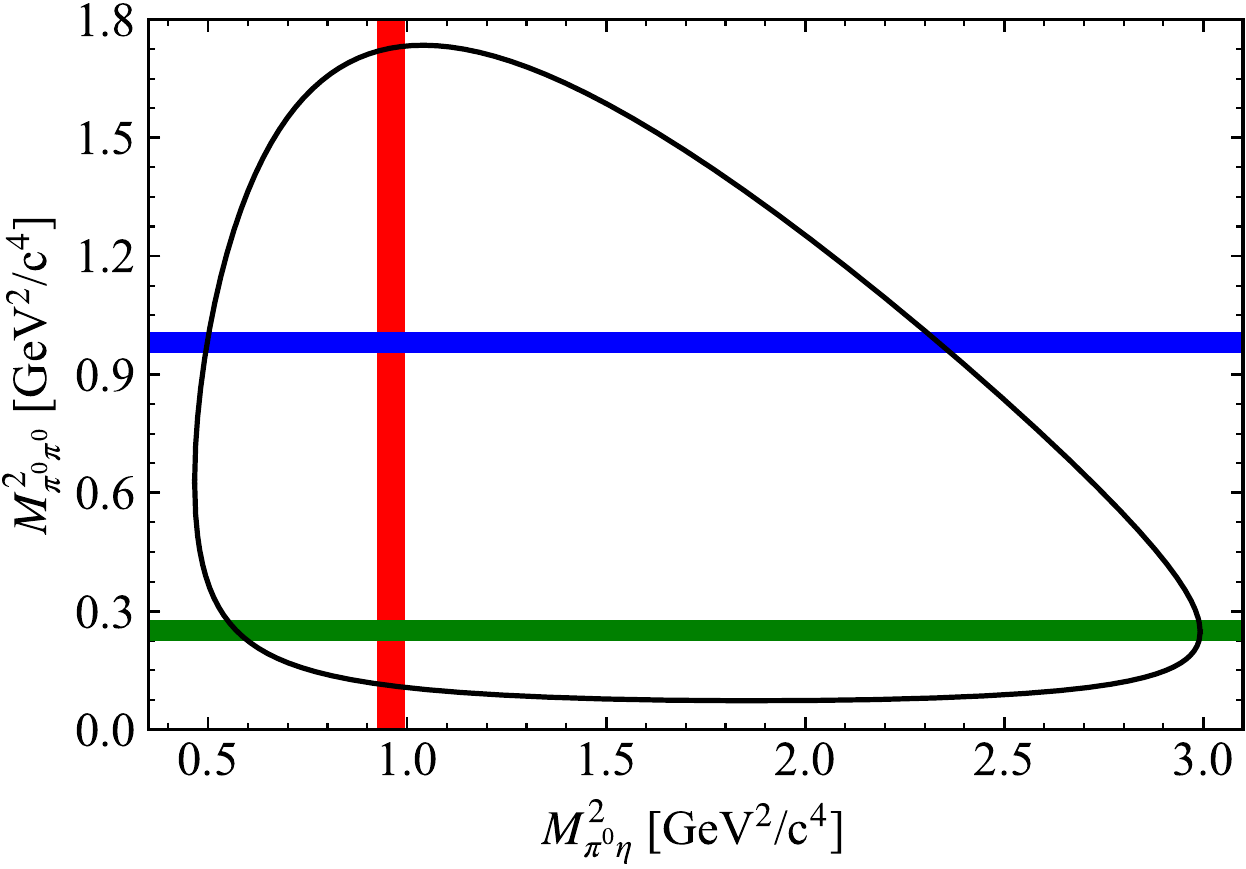} 
	\caption{\footnotesize The Dalitz plot of the $D^{0} \rightarrow \pi^{0}\pi^{0}\eta$ decay.}
	\label{fig:2Dalitzplot}
\end{figure}

\begin{figure}[htbp]
	\begin{subfigure}{0.45\textwidth}
		\centering
		\includegraphics[width=1\linewidth]{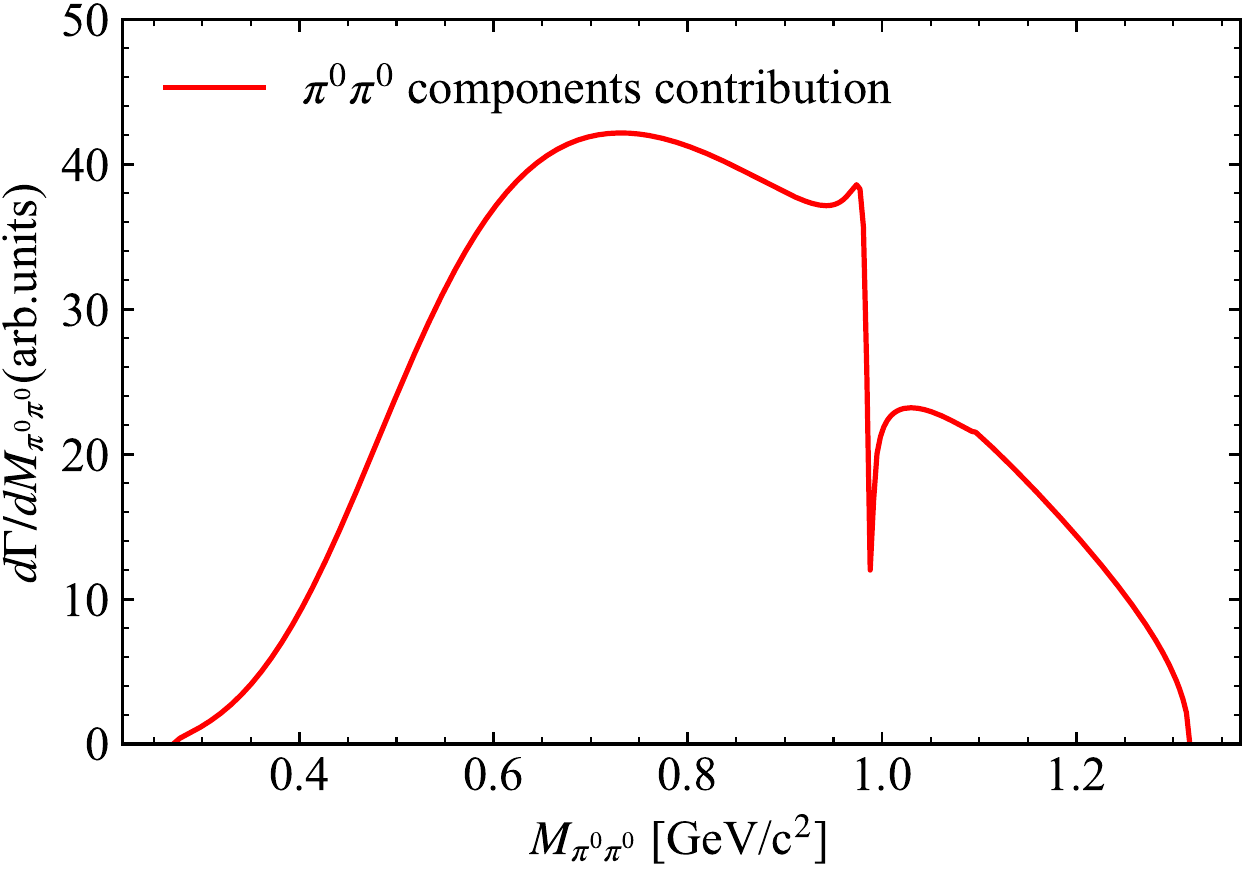} 
		\caption{\footnotesize}
		\label{fig:2dGamma12I0}
	\end{subfigure}
	\quad
	\quad
	\begin{subfigure}{0.45\textwidth}  
		\centering 
		\includegraphics[width=1\linewidth]{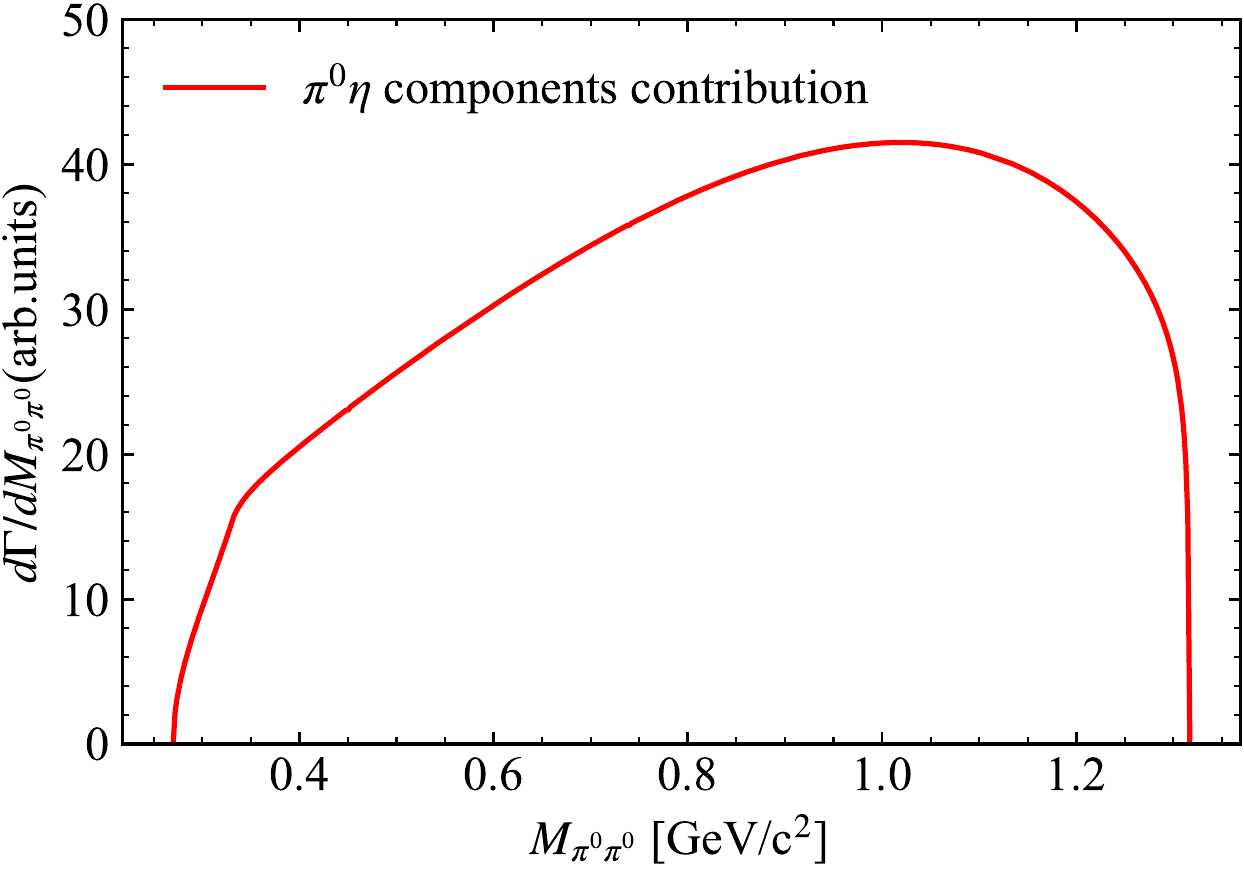} 
		\caption{\footnotesize}
		\label{fig:2dGamma12I1}  
	\end{subfigure}	
	\quad
	\quad
	\begin{subfigure}{0.45\textwidth}  
		\centering 
		\includegraphics[width=1\linewidth]{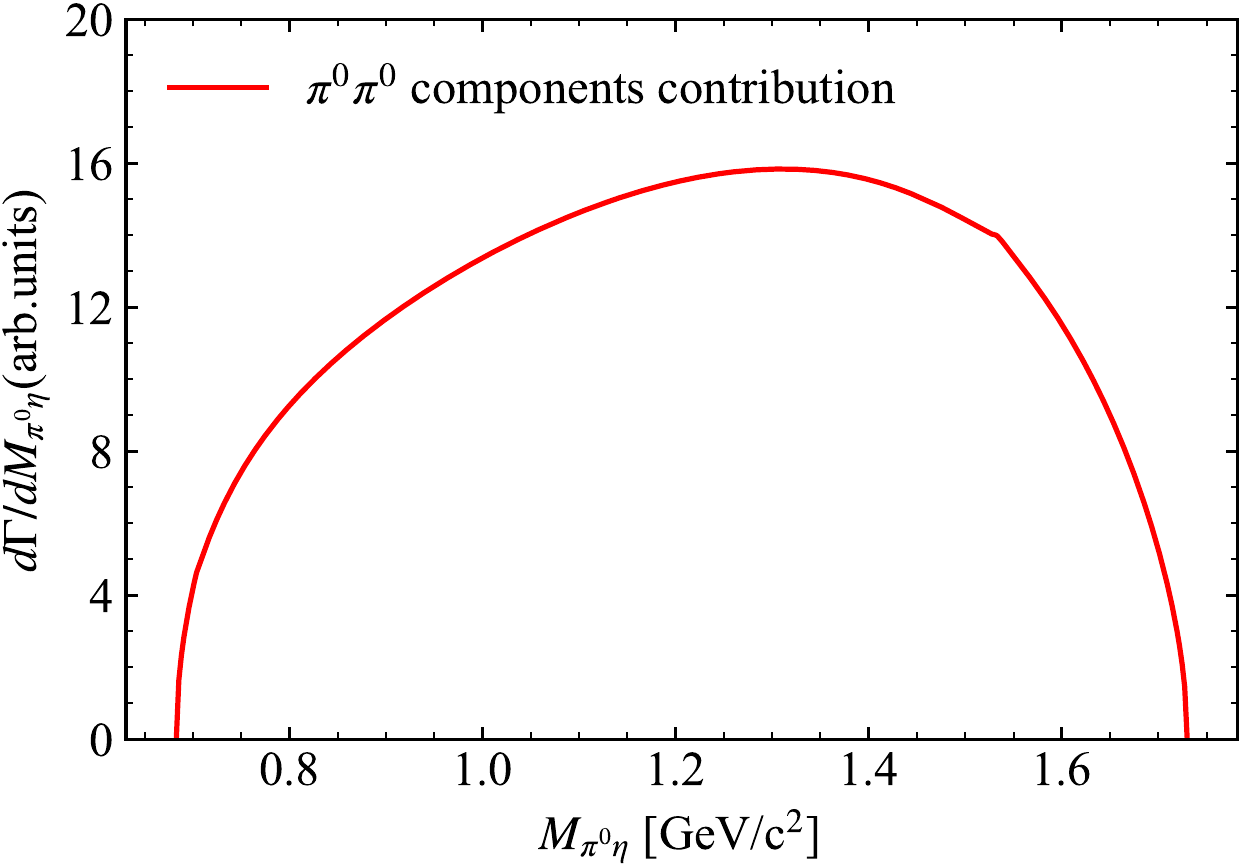} 
		\caption{\footnotesize}
		\label{fig:2dGamma13I0}  
	\end{subfigure}	
	\quad
	\quad
	\begin{subfigure}{0.45\textwidth}  
		\centering 
		\includegraphics[width=1\linewidth]{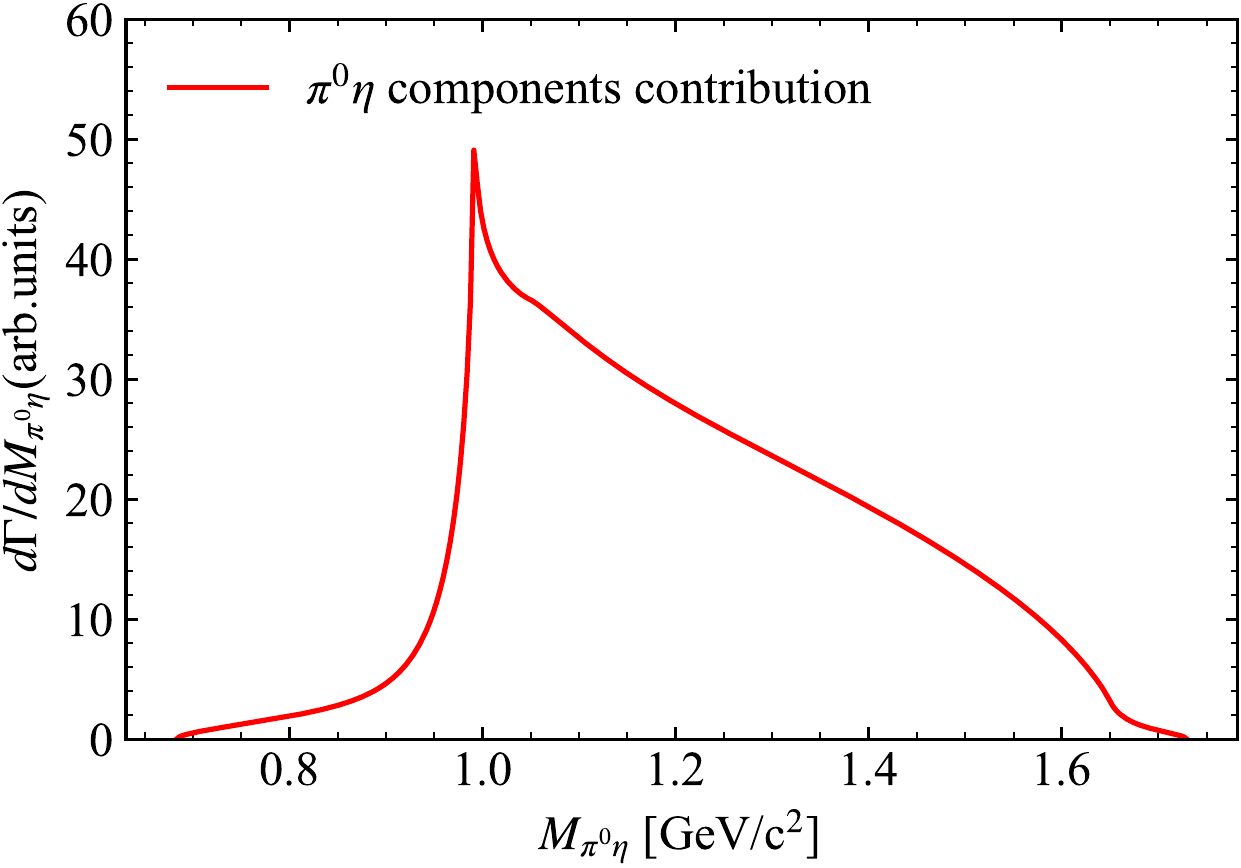} 
		\caption{\footnotesize}
		\label{fig:2dGamma13I1}  
	\end{subfigure}	
	\caption{\footnotesize The $\pi^{0}\pi^{0}$ (a) and $\pi^{0}\eta$ (b) components contributed to the $\pi^{0}\pi^{0}$ invariant mass distribution of the $D^{0} \rightarrow \pi^{0}\pi^{0}\eta$ decay, the $\pi^{0}\pi^{0}$ (c) and $\pi^{0}\eta$ (d) components contributed to the $\pi^{0}\eta$ invariant mass distribution of the $D^{0} \rightarrow \pi^{0}\pi^{0}\eta$ decay.}
	\label{fig:2dGammaI0I1}
\end{figure}

In Fig. \ref{fig:2dGamma}, we show the $\pi^{0}\pi^{0}$ invariant mass distribution in the $D^{0} \rightarrow \pi^{0}\pi^{0}\eta$ decay in sub-figure (a), and the one of $\pi^{0}\eta$ in sub-figure (b). One can see the $a_{0}(980)$ signal in the $\pi^{0}\eta$ invariant mass spectrum, and the $f_{0}(980)$ and $f_{0}(500)$ signal in the one of $\pi^{0}\pi^{0}$ components. Note that, both the $\pi^{0}\pi^{0}$ and $\pi^{0}\eta$ components have significant contributions in the final sate interactions of the $D^{0} \rightarrow \pi^{0}\pi^{0}\eta$ decay. This is different from the case of the $D^{0} \rightarrow \pi^{0}\pi^{0}\pi^{0}$ decay, where there is only $\pi^{0}\pi^{0}$ invariant mass distribution, and the case of the $D^{0} \rightarrow \pi^{0}\eta\eta$, where the interaction of $\pi^{0}\eta$ components is little affected by the ones of $\eta\eta$. The Dalitz plot for the $D^{0} \rightarrow \pi^{0}\pi^{0}\eta$ is shown in Fig. \ref{fig:2Dalitzplot}, where the red solid line (vertical one) is the position of $a_{0}(980)$ state, the blue solid line (upper horizontal one) the position of $f_{0}(980)$ and the green solid line (lower horizontal one) the $f_{0}(500)$ state, where the PDG value for the masses of each particle \cite{Zyla:2020pdg} used in the plot. The $a_{0}(980)$ contributes in the region of $0.15<M_{\pi^{0}\pi^{0}}^{2}<1.7$ GeV$^{2}$/c$^{4}$ of the $\pi^{0}\pi^{0}$ invariant mass distribution, the $f_{0}(980)$ contributes in the region of $0.5<M_{\pi^{0}\pi^{0}}^{2}<2.4$ GeV$^{2}$/c$^{4}$ and the $f_{0}(500)$ contributes in the region of $0.6<M_{\pi^{0}\pi^{0}}^{2}<3.0$ GeV$^{2}$/c$^{4}$ of the $\pi^{0}\eta$ invariant mass distribution. Then we analyze the contributions of the $\pi^{0}\pi^{0}$ and $\pi^{0}\eta$ components to the $\pi^{0}\pi^{0}$ invariant mass spectrum, which is shown in Fig. \ref{fig:2dGammaI0I1}. The broad peak of the $f_{0}(500)$ is obvious in Fig. \ref{fig:2dGamma12I0}, which should be a contribution from the transition $\pi^{+}\pi^{-} \rightarrow\pi^{0}\pi^{0}$ in Eq. (\ref{eq:amplitudes2}), and the small peak near the $K\bar K$ threshold is the state $f_{0}(980)$, which should be a contribution from the transition $K^{+}K^{-} \rightarrow\pi^{0}\pi^{0}$ in Eq. (\ref{eq:amplitudes2}). In Fig. \ref{fig:2dGamma12I1}, there is no obvious peak structure in the invariant mass spectrum of $\pi^{0}\pi^{0}$ from the contribution of $\pi^{0}\eta$ parts. Thus, we can confirm that in Fig. \ref{fig:2dGamma12}, the low energy region is the $f_{0}(500)$, and the peak near the $K\bar K$ threshold is the $f_{0}(980)$, which is enhanced by interference effect with the contribution of the $\pi^{0}\eta$ components. Then from the results shown in Figs. \ref{fig:2dGamma13I0} and \ref{fig:2dGamma13I1}, we know the peak structure near the $K\bar K$ threshold in Fig. \ref{fig:2dGamma13} is the $a_{0}(980)$ resonance with no $f_{0}(980)$ contribution.

Then we make some predictions for the ratios of branching fractions in different decay processes. In our theoretical model, the parameter of the production vertex $V_{P}$ in Eqs. (\ref{eq:Ha2}-\ref{eq:Hd2}) is unknown, see more discussion in Ref. \cite{Ahmed:2020qkv}. Thus we calculate the ratios of branching fractions for different decay channels as follows, where the unknown production vertex $V_{P}$ can be cancelled. Therefore, the results for these ratios are independent with parameter $V_{P}$ and more reliable. By integrating the invariant mass variables in the $D^{0} \rightarrow \pi^{0}\pi^{0}\pi^{0}$ and $D^{0} \rightarrow \pi^{0}\eta\eta$ decays over the invariant mass distribution, we find

\begin{equation} 
	\begin{aligned}
		\frac{\mathcal{B}[D^{0} \rightarrow f_{0}(980) \pi^{0}, f_{0}(980) \rightarrow \pi^{0}\pi^{0}]}{\mathcal{B}[D^{0} \rightarrow a_{0}(980) \eta, a_{0}(980) \rightarrow \pi^{0}\eta]} = 1.01 _{-0.10}^{+0.10},
	\end{aligned}
	\label{eq:Ratios1}
\end{equation} 

\noindent
where the integral limits are $[2m_{\pi^{0}},1.2]$ GeV and $[m_{\pi^{0}}+m_{\eta},1.2]$ GeV for $D^{0} \rightarrow f_{0}(980) \pi^{0}$ and $D^{0} \rightarrow a_{0}(980) \eta$, respectively, with the uncertainties from the integrated upper limit $1.2 \pm 0.05 $ GeV. Analogously, we get

\begin{equation} 
	\begin{aligned}
		\frac{\mathcal{B}[D^{0} \rightarrow a_{0}(980) \pi^{0}, a_{0}(980) \rightarrow \pi^{0}\eta]}{\mathcal{B}[D^{0} \rightarrow a_{0}(980) \eta, a_{0}(980) \rightarrow \pi^{0}\eta]} = 1.87 _{-0.23}^{+0.22},
	\end{aligned}
	\label{eq:Ratios2}
\end{equation}

\begin{equation} 
	\begin{aligned}
		\frac{\mathcal{B}[D^{0} \rightarrow f_{0}(500) \eta, f_{0}(500) \rightarrow \pi^{0}\pi^{0}]}{\mathcal{B}[D^{0} \rightarrow a_{0}(980) \eta, a_{0}(980) \rightarrow \pi^{0}\eta]} = 3.50 _{-0.53}^{+0.54},
	\end{aligned}
	\label{eq:Ratios3}
\end{equation}

\begin{equation} 
	\begin{aligned}
		\frac{\mathcal{B}[D^{0} \rightarrow f_{0}(980) \eta, f_{0}(980) \rightarrow \pi^{0}\pi^{0}]}{\mathcal{B}[D^{0} \rightarrow a_{0}(980) \eta, a_{0}(980) \rightarrow \pi^{0}\eta]} = 2.55 _{-0.50}^{+0.48},
	\end{aligned}
	\label{eq:Ratios4}
\end{equation}

\noindent
where the integral limits and uncertainties for the $D^{0} \rightarrow a_{0}(980) \pi^{0}$ decay are the same as $D^{0} \rightarrow a_{0}(980) \eta$. For the decays of $D^{0} \rightarrow f_{0}(500) \eta$ and $D^{0} \rightarrow f_{0}(980) \eta$, the integral limits are $[2m_{\pi^{0}}, 0.9]$ GeV and $[0.9, 1.2]$ GeV, respectively, where the uncertainties are obtained from the integrated limit of $0.9 \pm 0.05 $ GeV, as done in the Ref \cite{Ahmed:2020qkv}. As one can see from the results in Eqs. (\ref{eq:Ratios1}-\ref{eq:Ratios4}), the branching fractions of these decay channels are at the same order of magnitude, where the different quantities are about from $1$ to $3.5$. The branching fractions of the decays $D^{0} \rightarrow \pi^{0}\pi^{0}\pi^{0}, \pi^{0}\pi^{0}\eta, \pi^{0}\eta\eta$ measured by BESIII Collaboration are also at the same order of magnitude \cite{BESIII:2018hui}, $\mathcal{B}\left(D^{0} \rightarrow \pi^{0}\pi^{0}\pi^{0}\right)=(2.0 \pm 0.4 \pm 0.3) \times 10^{-4}$, $\mathcal{B}\left(D^{0} \rightarrow \pi^{0}\pi^{0}\eta\right)=(3.8 \pm 1.1 \pm 0.7) \times 10^{-4}$ and $\mathcal{B}\left(D^{0} \rightarrow \pi^{0}\eta\eta\right)=(7.3 \pm 1.6 \pm 1.5) \times 10^{-4}$, which imply that the scalar resonances of $f_{0}(500)$, $f_{0}(980)$ or $a_{0}(980)$ are dominant in these $D^{0}$ meson decay processes. We hope that their contributions can be measured in the future experiments.

\section{Conclusions}
\label{sec:Conclusions}

In the present work, we make a theoretical study of the singly Cabibbo-suppressed processes of $D^{0} \rightarrow \pi^{0}\pi^{0}\pi^{0}, \pi^{0}\pi^{0}\eta, \pi^{0}\eta\eta$ by taking into account the final state interactions. We have presented the $\pi^{0}\pi^{0}$ and the $\pi^{0}\eta$ invariant mass distributions of these decay processes, where the scalar resonances $f_{0}(500)$, $f_{0}(980)$ and $a_{0}(980)$ are dynamically generated in $S$-wave interactions with the ChUA. The results are in a good agreement with the experimental data with only one parameter. Indeed, the final states $\pi^{0}\pi^{0}\pi^{0}$, $\pi^{0}\pi^{0}\eta$ and $\pi^{0}\eta\eta$ are not possibly produced at the tree level [see Eqs.(\ref{eq:amplitudes1}-\ref{eq:amplitudes3})], and all the contributions come from the rescattering of the two-body final state interactions. For the $D^{0} \rightarrow \pi^{0}\pi^{0}\pi^{0}$ decay, the dominant contribution comes from the $I=0$ resonance $f_{0}(980)$. For the $D^{0} \rightarrow \pi^{0}\pi^{0}\eta$ decay, the contributions come from the $I=0$ states $f_{0}(500)$ and $f_{0}(980)$ in the $\pi^{0}\pi^{0}$ components, and the $I=1$ state $a_{0}(980)$ in the ones of $\pi^{0}\eta$. For the $D^{0} \rightarrow \pi^{0}\eta\eta$ decay, the dominant contribution is the one of $I=1$ resonance $a_{0}(980)$. With the analysis of the invariant mass spectra and corresponding amplitudes, we find that the main components of $f_{0}(500)$ is the $\pi\pi$ and the dominant components of $f_{0}(980)$ is the $K\bar K$, which is consistent with the analysis of Ref. \cite{Ahmed:2020kmp}. These results indicate that these resonances as being dynamically generated from the final state interactions of pseudoscalar meson pairs. Moreover, we also calculate the ratios of corresponding branching fractions. Finally, we hope that our predicted $\pi^{0}\pi^{0}$ invariant mass distribution for the decay of $D^{0} \rightarrow \pi^{0}\pi^{0}\pi^{0}$, and $\pi^{0}\pi^{0}$, $\pi^{0}\eta$ invariant mass distributions for the decay of $D^{0} \rightarrow \pi^{0}\pi^{0}\eta$ can be measured by future experiments.\\

\section*{Acknowledgements}
We thank Prof. Eulogio Oset for careful reading the manuscript and valuable comments, and acknowledge Prof. N. N. Achasov for useful comments.

 \addcontentsline{toc}{section}{References}
\end{document}